\documentclass[aip, 
jcp, longbibliography, 
twocolumn, floatfix, superscriptaddress, reprint, 10pt]{revtex4-1}

\usepackage[dvipsnames]{xcolor}
\usepackage[english]{babel}
\usepackage[utf8]{inputenc}
\usepackage[paperwidth=210mm,paperheight=297mm,centering,hmargin=2cm,vmargin=2.5cm]{geometry}
\usepackage[utf8]{inputenc}
\usepackage{verbatim}
\usepackage{comment}
\usepackage{bm}
\usepackage{amsmath}
\usepackage{amsfonts}
\usepackage{natbib}
\usepackage{pdfpages}
\usepackage{graphics}
\usepackage{float}
\usepackage{hyperref}
\usepackage[capitalize]{cleveref}
\usepackage{multirow}
\usepackage{booktabs}
\usepackage{makecell}
\usepackage{outlines}
\usepackage{lipsum}
\usepackage{siunitx}
\usepackage[version=4]{mhchem}
\usepackage{ulem}

\usepackage{multirow}

\ifdefined\DIRACREP
\else
\newcommand{\DIRACREP}{}

\usepackage{physics}
\usepackage{xparse}
\ifdefined\COSMOMATHS
\else
\newcommand{\COSMOMATHS}{}
\newcommand{\D}[1]{\operatorname{d}{\!#1}\,}

\fi %

\NewDocumentCommand{\rep}{s d<| d|>}{%
\IfBooleanTF{#1}{
   \IfValueTF{#2}{
       \IfValueTF{#3}{\braket{#2}{#3}}{\bra{#2}}
       }{
       \IfValueTF{#3}{\ket{#3}}{}
       }
   }{
   \IfValueTF{#2}{
       \IfValueTF{#3}{\braket*{#2}{#3}}{\bra*{#2}}
       }{
       \IfValueTF{#3}{\ket*{#3}}{}
       }
   }
}

\NewDocumentCommand{\rbra}{sm}{\IfBooleanTF{#1}{\rep*<#2|}{\rep<#2|}}
\NewDocumentCommand{\rket}{sm}{\IfBooleanTF{#1}{\rep*|#2>}{\rep|#2>}}
\NewDocumentCommand{\rbraket}{smom}{
    \IfBooleanTF{#1}{
        \IfNoValueTF{#3}{\rep*<#2||#4>}{\rep*<#2|#3\rep*|#4>}
    }{
        \IfNoValueTF{#3}{\rep<#2||#4>}{\rep<#2|#3\rep|#4>}
    }
}

\NewDocumentCommand{\cg}{m m m}{\rep<#1; #2||#3>}

\NewDocumentCommand{\field}{o m e{_} e{^} o e{_} e{^}}{
\IfValueTF{#5}{\overline{
  #2\IfValueT{#3}{_#3}\IfValueT{#4}{^{\otimes #4}} %
  \otimes
  #5\IfValueT{#6}{_#6}\IfValueT{#7}{^{\otimes #7}} %
  \IfValueT{#1}{;#1}
}}{
  \IfValueTF{#4}{\overline{
     #2\IfValueT{#3}{_#3}\IfValueT{#4}{^{\otimes #4}}
     \IfValueT{#1}{;#1}
  }}
  {#2\IfValueT{#3}{_#3}}
}
}

\NewDocumentCommand{\frho}{o e{_} e{^}}{
\field[#1]{\rho}_{#2}^{#3}
}

\newcommand{\br}{\mathbf{r}}
\newcommand{\bx}{\mathbf{x}}
\newcommand{\bxhat}{\hat{\bx}}
\newcommand{\brhat}{\hat{\br}}

\newcommand{\e}{a}  %

\NewDocumentCommand{\ex}{e_}{
\IfValueTF{#1}{\e_{#1}\bx_{#1}}{\e\bx}
}  %

\NewDocumentCommand{\lm}{e_}{
\IfValueTF{#1}{l_{#1}m_{#1}}{lm}
}
\NewDocumentCommand{\nlm}{e_}{
\IfValueTF{#1}{n_{#1}\lm_{#1}}{n\lm}
}
\NewDocumentCommand{\enlm}{e_}{
\IfValueTF{#1}{\e_{#1}\nlm_{#1}}{\e\nlm}
}
\NewDocumentCommand{\en}{e_}{
\IfValueTF{#1}{\e_{#1}n_{#1}}{\e n}
}
\NewDocumentCommand{\nlk}{e_}{
\IfValueTF{#1}{n_{#1}l_{#1}k_{#1}}{nlk}
}
\NewDocumentCommand{\enlk}{e_}{
\IfValueTF{#1}{\e_{#1}\nlk_{#1}}{\e\nlk}
}
\NewDocumentCommand{\enl}{e_}{
\IfValueTF{#1}{\en_{#1}l_#1}{\en l}
}

\NewDocumentCommand{\nnl}{s}{
\IfBooleanTF{#1}{n_1 n_2 l}{n_1; n_2; l}
}
\NewDocumentCommand{\ennl}{s}{
\IfBooleanTF{#1}{\en_1 \en_2 l}{\en_1; \en_2; l}
}

\NewDocumentCommand{\gslm}{s}{
\IfBooleanTF{#1}{\sigma\lambda\mu}{\sigma;\lambda\mu}
}

\newcommand{\rcut}[0]{{r_\text{cut}} }

\fi %

\newcommand{\ntrain}{\ensuremath{n_\text{train}}}
\newcommand{\rev}[1]{#1}

\usepackage{pdfpages}
\usepackage{pgffor}
\makeatletter
\AtBeginDocument{\let\LS@rot\@undefined}
\makeatother
\begin{document}

\setcitestyle{super}

\title{A smooth basis for atomistic machine learning}
\author{Filippo Bigi}
\affiliation{Physical and Theoretical Chemistry Laboratory, South Parks Road, Oxford OX1 3QZ, UK}

\author{Kevin Huguenin-Dumittan}
\author{Michele Ceriotti}
\affiliation{Laboratory of Computational Science and Modeling, Institute of Materials, \'Ecole Polytechnique F\'ed\'erale de Lausanne, 1015 Lausanne, Switzerland}

\author{David E. Manolopoulos}
\affiliation{Physical and Theoretical Chemistry Laboratory, South Parks Road, Oxford OX1 3QZ, UK}

\onecolumngrid
\begin{abstract}

Machine learning frameworks based on correlations of interatomic positions begin with a discretized description of the density of other atoms in the neighbourhood of each atom in the system. Symmetry considerations support the use of spherical harmonics to expand the angular dependence of this density, but there is as yet no clear rationale to choose one radial basis over another. Here we investigate the basis that results from the solution of the Laplacian eigenvalue problem within a sphere around the atom of interest. 
\rev{ We show that this generates the a basis of controllable smoothness within the sphere (in the same sense as plane waves provide a basis with controllable smoothness for a problem with periodic boundaries), and that a tensor product of Laplacian eigenstates also provides a smooth basis for expanding any higher-order correlation of the atomic density within the appropriate hypersphere. }
We consider several unsupervised metrics of the quality of a basis for a given dataset, and show that the Laplacian eigenstate basis has a performance that is much better than some widely used basis sets and competitive with data-driven bases that numerically optimize each metric. \rev{Finally, we investigate the role of the basis in building models of the potential energy. In these tests, we find that a combination of the Laplacian eigenstate basis and target-oriented heuristics leads to equal or improved regression performance when compared to both heuristic and data-driven bases in the literature. We conclude that the smoothness of the basis functions is a key aspect of successful atomic density representations.}
\end{abstract}
\twocolumngrid

\maketitle

\section{Introduction}

Machine learning (ML) has gained increasing importance in the field of atomistic modeling during the last decade. Successful applications have involved both supervised learning (often in the form of regression of atomic-scale properties) and unsupervised learning (e.g.,  clustering of large compound/structure databases). Most ML methods, whether supervised or unsupervised, rely on representations of the atomic structures of interest. These representations are constructed as a set of numerical descriptors (or features) which act as the inputs to the ML model. Although several classes of descriptors are now available, their performance is often limited by the choice of basis that is used to project the atomic density correlations \cite{musi+21cr}. As a result, the identification of a suitable basis is a key step in building more effective descriptors and, ultimately, ML models. 

Desirable properties of atomic-scale descriptors include equivariance,  uniqueness, interpretability, and low computational cost\cite{onat+20jcp}. However, most of these are properties of broad classes of descriptors, which will not be our focus here. Instead we shall confine our attention to properties that are directly affected by the choice of basis, such as low dimensionality and high information content of the descriptor space, sensitivity to changes in the atomic configuration, and good performance on regression tasks.
The spherical harmonics are (almost\cite{shap16mms}) always used as the angular basis for atomic density expansions\cite{musi+21cr, bart+10prl, bart+13prb, wood-thom18jcp, drau19prb, niga+20jcp}. This is because of their symmetry properties with respect to rotation and inversion, which make it possible to generate features with the desired equivariant behavior. However, the best choice of the radial basis is still an open question, with a variety of different radial bases currently in use in different packages for atomic-scale representations and machine learning\cite{bart+10prl, bart-csan15ijqc, hima+20cpc, musi+21jcp, dusson_2022_atomic}. 

Radial bases have traditionally been regarded as a component of atomic density-based approaches with a very large design space, but perhaps not a significant effect on the quality of descriptors. As a result, the radial basis functions have typically been chosen for their simplicity and/or computational efficiency. However, these considerations become irrelevant when one uses cubic splines to evaluate the radial basis functions, which can then all be computed equally efficiently\cite{musi+21jcp}. This is true regardless of their complexity, and regardless of whether the basis is ``mollified'' as a result of Gaussian smearing of the atomic densities. The focus therefore shifts to the search for a radial basis with optimal properties, regardless of its complexity. 

In the present paper, we shall show that the radial basis, and, perhaps even more importantly, how it is truncated, can have a significant impact on the quality of the resulting descriptors. Following a comparison with several other radial bases, ranging from traditional bases to more recent data-driven contractions\cite{gosc+21jcp}, we identify the Laplacian eigenstates within a sphere around each atom\cite{klic+20arxiv, onthefly} as a particularly elegant basis that results in density expansion coefficients and structural descriptors exhibiting all of the desirable properties listed above. 
We then proceed, motivated by the recent focus on systematic body-ordered expansions\cite{shap16mms, drau19prb, niga+20jcp}, to investigate the behavior of this and other bases in the context of the higher-order equivariant features that arise in the symmetry-adapted expansion of many-body density correlations.
\rev{We find that the LE basis, and the truncation strategy we propose, can be successfully combined with several of the heuristic arguments that are commonly applied to the construction of machine-learning potentials\cite{behl11jcp,huan-vonl16jcp,drau19prb,novi+21mlst}.}

\section{Theory}

\subsection{Density expansion and density correlations}

Machine learning of properties of local atomic environments begins by constructing an equivariant discretization of the neighbor density around each atom $i$ in a structure $A$
\begin{equation}
\rho_i(\bx) = \sum_{j\in A} g(\bx -\br_{ji}) \approx \sum_{nlm}  \psi_{nlm}(\bx)c^i_{nlm}, \label{eq:rhoi-expansion}
\end{equation}
where $\br_{ji}=\br_j-\br_i$ is the vector between the central atom and its $j$-th neighbor, $g(\bx)$ is a localized density function (usually a Dirac delta function or a Gaussian), the $c^i_{nlm}$ are the discrete expansion coefficients for the $i$-th environment, and the $\psi_{nlm}(\bx)$ are orthonormal basis functions. 
A number of different basis sets have been proposed for this purpose,  the vast majority of which use spherical harmonics to represent the angular dependence of the density in conjunction with various radial basis functions $R_{nl}(x)$:
\begin{equation}
\psi_{nlm}(x\bxhat) = R_{nl}(x) Y_l^m(\bxhat).\label{eq:psinlm}
\end{equation}
We use real-valued spherical harmonics $Y_l^m(\bxhat)$ so that the expansion coefficients $c^i_{nlm}$ are real.

The same expressions can be written in a form that borrows from the Dirac bra-ket notation of quantum mechanics,
\begin{equation}
\rep<\bx||\rho_i> = \sum_{j\in A} \rep<\bx||\br_{ji}; g> \approx \sum_{nlm} \rep<\bx||nlm> \rep<nlm||\rho_i>,
\end{equation}
in which the terms are in one-to-one correspondence with those in Eq.~\eqref{eq:rhoi-expansion}.
The expansion coefficients can be written as 
\begin{multline}
c^i_{nlm}=\sum_{j\in A} \int\D{\bx} R_{nl}(x)Y_l^m(\bxhat) g(\bx-\br_{ji}) \\
=\sum_{j\in A} \rep<nlm||\br_{ji}; g> = \rep<nlm||\rho_i>.\label{eq:expan-coeffs}\phantom{xxxxxxx.}
\end{multline}
A pedagogic introduction to the use of this notation is given in Section 3.1 of Ref.\citenum{musi+21cr}.

We will employ both the functional and the bra-ket notation, using the former when discussing specific basis functions and the latter when manipulating expansion coefficients in a way that does not depend on the choice of the basis. In particular, we will use bra-ket notation to express the equivariant coefficients that discretize the $\nu$-point correlations of the neighbor density as $\rep<q ||\frho[\sigma; \lambda\mu]_{i}^{\nu}>$. This notation singles out the components of the $\nu$-point density correlation $\rho_i^{\otimes\nu}$ that acquire a phase of $(-1)^\lambda\sigma$ under inversion and transform under rotation as $Y^\mu_\lambda$. The indices in the ket thus identify the symmetry properties of the features, while the $q$ in the bra is in general a composite index that we use to enumerate concisely all features associated with a given representation that share the same symmetry. 

As discussed in Ref.~\citenum{niga+20jcp}, these equivariant features can be built by starting from the density expansion coefficients that define the $\nu=1$ equivariants
\begin{equation}
\rep<n||\frho[1; \lambda\mu]_i^1> = \rep<n\lambda\mu||\rho_i>
\end{equation}
and iterating an angular-momentum sum rule
\begin{multline}
\rep<q l_2; n l_1  ||\frho[\sigma; \lambda\mu]_i^{\nu+1}>=\sum_{m_1 m_2}\rep<n||\frho[1; l_1 m_1]_i^1>\\\times   \rep<q|| \frho[\sigma (-1)^{l_1+l_2+\lambda}; l_2 m_2 ]_i^\nu> \!\cg{l_1 m_1}{l_2 m_2}{\lambda \mu },
\label{eq:gen-cg-iter}
\end{multline}
which uses modified Clebsch-Gordan (CG) coefficients $\cg{l_1 m_1}{l_2 m_2}{\lambda \mu }$ that are appropriate for the combination of real spherical harmonics.

The atom-centered density correlations (ACDC) that can be constructed from the $\rep<q ||\frho[\sigma; \lambda\mu]_{i}^{\nu}>$ encompass many of the representations that have been proposed over the past decade, including atom-centered symmetry functions\cite{behl-parr07prl} and the SOAP power spectrum\cite{bart+10prl} (both equivalent to $\nu=2$ invariants), the SOAP bispectrum\cite{wood-thom18jcp} (equivalent to $\nu=3$ invariants), $\lambda$-SOAP features ($\nu=2$ equivariants), and more recent systematic expansions such as the atomic cluster expansion\cite{drau19prb}, in which the invariant correlations are used as a basis to obtain interatomic potentials as linear fits. Moment tensor potentials\cite{shap16mms}, achieve a similar goal with a formalism based on Cartesian rather than spherical coordinates. 
Equivariant neural networks\cite{kondor2018covariant,thomas2018tensor, kondor2018clebsch,anderson2019cormorant,batzner2021se3,qiao2021unite} can also be shown to be very closely related to this construction\cite{niga+22jcp2}. 
Given that the number of equivariant features generated by iterating~\eqref{eq:gen-cg-iter} grows as the size of the density expansion basis to the power $\nu$,\footnote{Angular momentum theory provides rules to eliminate some redundant terms \cite{niga+20jcp}, but does not eliminate the exponential scaling}  the choice of a concise yet informative discretization is crucial to keep the computational cost under control.

\subsection{Laplacian eigenstates}\label{sec:LEbasis}

There are good reasons why the spherical harmonics are used to expand the angular dependence of the density. They are the basis functions of the irreducible representations of SO(3), and they are the eigenstates of the kinetic energy operator on the surface of a sphere (the Legendrian). 
Because they are irreducible representations of SO(3), the set of $(l_{\rm max}+1)^2$ spherical harmonics $Y_{l}^m(\hat\bx)$ with $l\le l_{\rm max}$ provides an equivariant representation of the density that treats all rotated densities on an equal footing, and because they are the Legendrian eigenstates with the smallest eigenvalues, this is the smoothest set of $(l_{\rm max}+1)^2$ orthonormal basis functions one can construct within a sphere, \rev{in the sense that they have the smallest mean square gradients over the sphere [see Eq.~\eqref{eq:rq}].}

All of the direct product bases of spherical harmonics and radial basis functions that have been proposed in the past lead to equivariant representations of the density. However, none of these previously proposed basis sets is \rev{is explicitly built to optimize some quantitative measure of its smoothness. }
Optimum smoothness is achieved by choosing the combinations of spherical harmonics and radial basis functions to be the lowest energy eigenstates of the kinetic energy operator {\em within} the sphere. This ensures that the combined radial and angular basis respects the uniformity of three dimensional space within the sphere in the same way as the spherical harmonic basis respects the isotropy of the two dimensional surface of the sphere.

For a sphere of radius $a$, the resulting Laplacian eigenstate (LE) basis functions are the solutions with the lowest eigenvalues of the equation
\begin{equation}
-\nabla^2\psi_{nlm}(\bx) = E_{nl}\psi_{nlm}(\bx), 
\label{SE}
\end{equation}
subject to the boundary condition
\begin{equation}
\psi_{nlm}(\bx) = 0\quad \hbox{for} \quad x=|\bx| = a
\label{Dirichlet}
\end{equation}
and the orthonormality condition
\begin{equation}
\int_{x<a}  \psi_{n'l'm'}(\bx)\psi_{nlm}(\bx)\,{\rm d}\bx = \delta_{n'n}\delta_{l'l}\delta_{m'm}.
\end{equation}
The solutions are given by Eq.~\eqref{eq:psinlm}, with radial functions $R_{nl}(x)$ whose explicit form is given along with that of the eigenvalues $E_{nl}$ in Appendix~\ref{app:le-details}. 

\begin{figure}[b]
\centering
\includegraphics[width=\linewidth]{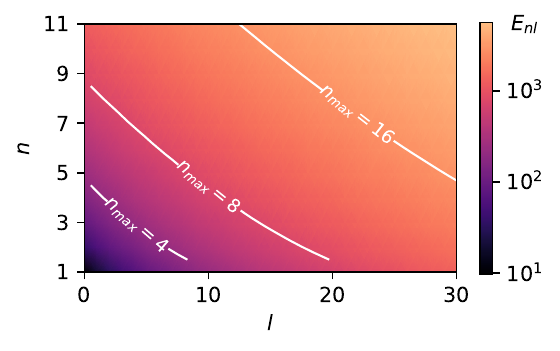}
\caption{A heat map of the Laplacian eigenvalue $E_{nl}$ in units of $1/a^2$, showing the regions of the $n,l$ plane that are included in the LE bases with $E_{\rm max}=E_{n_{\rm max},0}=(n_{\rm max}\pi/a)^2$ for $n_{\rm max}$ = 4, 8, and 16.}
\label{fig:nmaxl}
\end{figure}

The connection between small eigenvalues of $-\nabla^2$ and smooth eigenfunctions can be understood by considering the Dirichlet problem applied to the 3D sphere. As solutions of the more general Dirichlet problem in a region $\Omega$ of $N$-dimensional real space $\bx^N$, the eigenstates of $-\nabla^2$ have the following property: each successive eigenstate $u(\bx)$, starting from the one with the lowest eigenvalue, minimises the Rayleigh quotient
\begin{equation}
Q  = \left.\displaystyle{\int_{\Omega} \left|\nabla u(\bx)\right|^2\,{\rm d}\bx}\middle/\displaystyle{\int_{\Omega} \left|u(\bx)\right|^2\,{\rm d}\bx}\right.
\label{eq:rq}
\end{equation}
subject to the constraint that $u(\bx)$ is zero on the boundary of $\Omega$  and orthogonal to all eigenstates with smaller eigenvalues. If we take $Q$ as a measure of smoothness, the LE basis truncated via the eigenvalue criterion is thus the smoothest possible basis within the region $\Omega$, which in our case is the 3D sphere, and the eigenvalue $E_{nl}$ quantifies the smoothness of each solution. Truncating the LE basis by retaining the functions with $E_{nl}\le E_{\rm max}$ is analogous to truncating the spherical harmonic basis by retaining the functions with $l\le l_{\rm max}$, and it reduces the specification of the basis to a single parameter. The constraint that $E_{nl}\le E_{\rm max}$ implies that fewer spherical harmonics will be retained in the basis as $n$ increases, so it will contain fewer functions than a direct product basis with the same values of $n_{\rm max}$ and $l_{\rm max}$. This is illustrated in Fig.~\ref{fig:nmaxl}, which shows the regions of the $n,l$ plane that are included in the LE bases with $E_{\rm max}=E_{n_{\rm max}0}$ for various values of $n_{\rm max}$.

\rev{
\subsection{Basis-set smoothness}

Smoothness is a desirable feature of machine-learning schemes because it results in a well-behaved interpolation of machine-learned properties to configurations that are not present in the training set. 
This is an established concept in the atomistic machine learning literature, where smoothness is often enforced through the regularizing effect of including penalty terms in the loss. 
Smoothing strategies range from the naive Gaussian prior of kernel ridge regression (KRR) to more sophisticated regularization schemes such as that of Ref. \citenum{vand+20mlst}. The Gaussian smoothing of atomic densities first introduced in SOAP\cite{bart+13prb} can also be considered as a regularization. 
For models involving a basis set representation of the atom density, the regularity of the approximation also depends on the smoothness of the basis functions.  Here we intend to show how using a quantifiable smoothness criterion to guide both the choice of the basis set and the truncation of higher-order density correlation features is beneficial to the performance of atomistic ML models.

The Rayleigh quotient in Eq.~\eqref{eq:rq} is clearly just one of many possible measures of the regularity of a function: one could take its mean square curvature, its Lipschitz constant, etc. 
Perhaps the best way to motivate our choice is to note that the solutions of the Dirichlet problem with periodic boundary conditions are plane waves. The notion that a truncated Fourier expansion is a natural way to enforce regularity is well established, and indeed in plane wave electronic structure codes it is customary to express the basis set truncation in terms of a kinetic energy cutoff. 
The LE construction yields an analogous basis that is adapted to the symmetry and the boundary conditions of the present problem. This idea has already been exploited in atomistic machine learning: Ref.~\citenum{onthefly} and the existing DimeNet framework~\cite{klic+20arxiv} already use spherical Bessel functions to evaluate radial descriptors, the latter arguing that their bounded maximum frequencies confer both smoothness and stability on the resulting model predictions.
}

\subsection{Unsupervised measures of  basis-set quality}\label{sec:metrics}

We shall now describe three different metrics that can be used to assess the quality of basis sets such as the LE basis introduced in Sec.~II.B: a residual variance\cite{gosc+21jcp} metric $\ell_X$, a residual Jacobian variance metric $\ell_J$, and a Jacobian condition number\cite{pars+21mlst} metric $\ell_{CN}$. For the first two of these, it is possible to construct a basis of any desired size that, for a given dataset, is optimal with respect to the metric, and so provides a gold standard against which other bases can be compared. We shall explain how to do this in Sec.~II.D. We do not yet know how to construct a finite basis that is optimal for the Jacobian condition number metric, but we do at least know the value that the metric takes in the limit of a complete basis\cite{pozd+21ore}, so we can use that as the gold standard instead.

\subsubsection{Residual variance}

Since the expansion in Eq.~\eqref{eq:rhoi-expansion} aims to discretize the $i$-centered neighbor density, a natural criterion for the quality of a basis is how well it minimizes the error in the approximation of $\rho_i(\bx)$. For a given dataset, this can be defined as the normalised $L^2$ residual 
\begin{equation}
\ell_X = {\displaystyle{
\sum_i \int \D{\bx}\,\Bigl|\rho_i(\bx) - \sum_q c_q^i \psi_q(\bx)\Bigr|^2}\over\displaystyle{\sum_i \int {\rm d}\bx\,\rho_i(\bx)^2}},
\end{equation}
where $q$ is a shorthand for the $nlm$ basis indices, and $i$ runs over all atomic environments in the dataset. 
If the discrete basis is orthonormal, then one can compute the numerator in this expression as the difference between the mean $L^2$ norm of the neighbor density and the norm of the coefficients $c_q^i=\rep<q||\rho_i>$, 
\begin{equation}
\ell_{X} = {\displaystyle{\sum_i \left[\int \D{\bx} \rep<\bx||\rho_i>^2-  \sum_q \rep<q||\rho_i>^2\right]}\over \displaystyle{\sum_i \int \D{\bx} \rep<\bx||\rho_i>^2}},
\label{eq:variance}
\end{equation}
or equivalently as the difference between the norm of the coefficients in a complete basis and those in a truncated discretization. The numerator in Eq.~\eqref{eq:variance} is not exactly a difference between two variances, because the density and the coefficients are usually not centered to have zero mean over the data set, but we shall nevertheless call $\ell_X$ a residual variance metric following Ref.~\onlinecite{gosc+21jcp}.

\subsubsection{Residual Jacobian variance}

The Jacobian of the features quantifies their sensitivity to atomic displacements. It can be written in terms of the derivatives of the continuous atom density as 
\begin{multline}
J_{\bx,i j\alpha} = \frac{\partial \rep<\bx||\rho_i>}{\partial r_{j\alpha}} \equiv \rep<\bx||\partial_{j\alpha}\rho_i>=
\sum_{j'} \frac{\partial \rep<\bx||\br_{j'i}; g>}{\partial r_{j\alpha}}\\
=-\frac{\partial g(\bx-\br_{ji})}{\partial x_\alpha}
\equiv \rep<\bx||\br_{ji}; \partial_{\alpha} g>,\phantom{xxxx.}
\end{multline}
where $\alpha$ runs over the Cartesian axes, and in terms of discrete basis states as 
\begin{equation}
J_{q,ij\alpha} = \frac{\partial \rep<q||\rho_i>}{\partial r_{j\alpha}}
\equiv \rep<q||\partial_{j\alpha}\rho_i>= \rep<q||\br_{ji}; \partial_{\alpha}g>.
\end{equation}
The Jacobian arises in the calculation of derivatives of ML models (e.g. forces), and it has been used in the past to assess the quality of atomistic representations -- both directly\cite{pozd+21ore} and via the sensitivity matrix\cite{onat+20jcp} $\mathbf{J}_i^T\mathbf{J}_i$.
In the present context, it is natural to define a residual Jacobian variance metric $\ell_J$ for this purpose by analogy with Eq.~\eqref{eq:variance},
\begin{equation}
\!\!\!\ell_{J} = {\displaystyle{\sum_{ij\alpha} \left[\int \D{\bx} \rep<\bx||\partial_{j\alpha}\rho_i>^2 -\sum_q \rep<q||\partial_{j\alpha} \rho_i>^2 \right]}\over\displaystyle{\sum_{ij\alpha} \int \D\bx \rep<\bx||\partial_{j\alpha}\rho_i>^2}}, \label{eq:Jacobianvariance}
\end{equation}
in which the numerator can again be written equivalently as the difference between the norm of the coefficients $\rep<q||\partial_{j\alpha}\rho_i>$ in a complete basis and those in a truncated discretization.

\subsubsection{Jacobian condition number}

We shall base our last metric on the \emph{condition number} of the Jacobian, that is the ratio between the largest and smallest non-zero singular values of $\mathbf{J}_i$. As discussed in Ref.~\citenum{pozd+21ore}, in the limit of a sharp Gaussian smearing $\sigma$ and of a complete basis, all the non-zero singular values of $\mathbf{J}_i$ take the same value, because the elements of $\mathbf{J}_i^T\mathbf{J}_i$ become
\begin{multline}
\int \D{\bx} \rep<\partial_{j\alpha}\rho_i||\bx>\rep<\bx||\partial_{j'\alpha'}\rho_i> = \\  
\int \D{\bx} \rep<\br_{ji}; \partial_{\alpha}g||\bx> \rep<\bx||\br_{j'i}; \partial_{\alpha'}g> \underset{\sigma\rightarrow0}{\sim} 
\delta_{\alpha\alpha'}\delta_{jj'} \frac{\sqrt{\pi}^3\sigma}{2},
\end{multline}
and so the condition number tends to one. An appropriate Jacobian condition number metric is therefore
\begin{equation}
\ell_{CN} = \frac{1}{\ntrain}\sum_i\frac{s_\text{max} (\mathbf{J}_i)}{s_\text{min}(\mathbf{J}_i)} -1,
\end{equation}
where $\ntrain$ is the number of $i$-centered environments in the training set and $s_{\rm max}(\mathbf{J}_i)$ and $s_{\rm min}(\mathbf{J}_i)$ are the largest and smallest non-zero singular values of $\mathbf{J}_i$ (the square roots of the largest and smallest eigenvalues of $\mathbf{J}_i^T\mathbf{J}_i$).

\rev{From a numerical perspective, condition numbers in the hundreds or thousands are perfectly manageable with floating point arithmetic. However, we would argue that it is beneficial to approach the limit  $\ell_{CN}\rightarrow 0$,} because this implies that the discretized expansion coefficients are equally sensitive to all types of distortions; i.e., that the basis set does not artificially make the descriptors more sensitive to some deformations than to others.
\rev{ Indeed, it has recently been shown that numerical instabilities that result in moderate values of the Jacobian condition number can affect substantially the performance of ML models\cite{pars-goed22jcp,pozd+22jcp}. }
For the $\ell_{CN}$ metric to be meaningful, there must be at least three times more functions in the basis set than the maximum number of neighbors of any atom $i$ in the data set, so that all the sensitivity matrices $\mathbf{J}_i^T\mathbf{J}_i$ have full rank. 
Furthermore, the cutoff that is applied in defining the density must be sharp (so that all retained neighbors $j$ of atom $i$ are fully inside the cutoff), and no feature tuning through a radial scaling of the neighbor contributions\cite{huan-vonl16jcp,will+18pccp} can be applied. 
\rev{Incidentally, the fact that the modulating the feature sensitivity in a physically-motivated manner has a significant positive impact on model performance reinforces the notion that it is better to avoid the unpredictable, noisy modulation of the sensitivity that is caused by the choice of a non-smooth basis.}

\subsection{Metric-optimized bases}

The logical next step after having defined metrics to assess the quality of a basis is to look for a construction that yields a basis that is optimal with respect to each metric for a given data set. For the residual variance metric $\ell_X$, this step has already been taken in Ref.~\citenum{gosc+21jcp}, so here we only sketch the main ideas. 

Starting from a basis set expansion in a large set of primitive basis functions $\{\ket{p}\}$, one considers an orthogonal contraction of the expansion coefficients
\begin{equation}
\rep<q||\rho_i> = \sum_{p} U_{qp} \rep<p||\rho_i>,
\label{eq:orthogonal-projection}
\end{equation}
in which the orthogonal matrix $\mathbf{U}$ is chosen to diagonalize the real symmetric \lq\lq covariance" matrix\cite{gosc+21jcp} 
\begin{equation}
C_{pp'}= \sum_i  \rep<p||\rho_i>  \rep<\rho_i||p'>,
\end{equation}
such that
\begin{equation}
\sum_{pp'} U_{qp}C_{pp'}U_{q'p'} = \gamma_q\,\delta_{qq'}.
\end{equation}
Then since
\begin{equation}
\sum_{iq} \rep<q||\rho_i>^2 = \sum_{q} \gamma_q,
\end{equation}
it is clear that the functions $\ket{q}$ with the largest eigenvalues $\gamma_q$ provide an optimum contracted basis for the minimization of $\ell_X$ in Eq.~\eqref{eq:variance}.

Symmetry considerations imply that for a randomly-oriented data set the cross-correlations between features with different SO(3) character must be zero, and so in practice one computes separate blocks of the covariance matrix for each equivariant component
\begin{equation}
C^{\lambda}_{pp'}= \sum_{i\mu}  \rep<p||\frho[1; \lambda\mu]_i^1>  \rep<p'||\frho[1; \lambda\mu]_i^1>.
\label{eq:Clambda}
\end{equation}
As discussed in Ref.~\citenum{gosc+21jcp}, it is possible to compute explicitly the radial functions associated with the optimal combinations, and to evaluate the expansion coefficients by approximating them with splines. Since the number of spherical harmonics increases with increasing $\lambda$, we retain the covariance eigenstates with the largest values of $\gamma_q/(2\lambda+1)$ so as to achieve the most efficient contraction to a given number of basis functions. Since it gives the optimum value of the metric $\ell_X$ for a given dataset, we shall refer to the resulting basis as the X-OPT basis set.

An optimal contracted basis for the Jacobian variance metric $\ell_J$ in Eq.~\eqref{eq:Jacobianvariance} can be constructed in the same way, by computing the Jacobian covariance matrix in the primitive basis
\begin{multline}
\!\!\!K^{\lambda}_{pp'} \!=\! \sum_{ij\alpha\mu} \rep<p||\field[1; \lambda\mu]{\partial_{j\alpha} \rho}_i^1>
\rep<p'||\field[1; \lambda\mu]{\partial_{j\alpha} \rho}_i^1>,
\label{eq:Klambda}
\end{multline}
diagonalizing it, and using the eigenvectors with the largest eigenvalues when weighted by $1/(2\lambda+1)$ to construct the contracted basis functions. We shall refer to the resulting basis as the J-OPT basis set. Both this and the X-OPT basis can be related to the maximization of a Rayleigh quotient akin to Eq.~\eqref{eq:rq}, as we discuss in Appendix~\ref{app:rayleigh} and in the SI. 

\subsection{Higher correlation orders}

Our discussion so far has focused on the discretization of the neighbor density, which when expressed using spherical harmonics for the angular part leads to the first-order atom-centered equivariant correlations $\rep<n||\frho[1;\lambda\mu]_i^1>$. 
High-accuracy models, however, require higher orders of ACDCs, either built explicitly or through an equivariant network architecture. 
One should therefore consider how the above ideas generalize to high-order equivariants. 

\subsubsection{Laplacian eigenstates} 

The solution of the Laplacian eigenvalue problem in Eq.~\eqref{SE} generates the smoothest possible basis in the 3D sphere $\Omega$. More generally, an analogous Dirichlet problem can be solved in the $\Omega^\nu$ hypersphere, where the density $\nu$-correlation $\rho_i^{\otimes \nu}$ is defined.
The associated Laplacian eigenvalue equation is
\begin{equation}
- \Big( \sum_{k=1}^\nu \mathbf{\nabla}_k^2 \Big) \, \Psi(\{\bx_k\}) = E \, \Psi(\{\bx_k\}).
\end{equation}
Since this is separable into $\nu$ distinct eigenvalue equations, one for each 3D sphere which constitutes the tensor product, the eigenstates are
\begin{equation}
\Psi_{\{n_k l_k m_k\}}(\{\bx_k\}) = \prod_{k=1}^\nu \psi_{n_k l_k m_k}(\bx_k),
\end{equation}
and the eigenvalues are
\begin{equation}
E_{\{n_kl_k\}} = \sum_{k=1}^\nu E_{n_k l_k},
\end{equation}
where $\psi_{nlm}(\bx)$ and $E_{nl}$ are the eigenstates and eigenvalues of Eq.~\eqref{SE}. Furthermore, when a tensor product of atomic densities $\rho_i^{\otimes \nu}(\{\bx_k\})$ is expanded in this basis, the resulting integral is also separable, giving
\begin{equation}
\begin{split}
\int_{\Omega^\nu} \Psi_{\{n_k l_k m_k\}}(\{\bx_k\}) \rho_i^{\otimes \nu}(\{\bx_k\}) \,\D\{\bx_k\} = \phantom{xxxxx}\\
\prod_{k=1}^\nu \int_\Omega \psi_{n_k l_k m_k}(\bx_k) \rho_i(\bx_k) \,  \D\bx_k  = \prod_{k=1}^\nu c^i_{n_k l_k m_k}, 
\end{split}
\end{equation}
where the $c^i_{nlm} = \rep<nlm||\rho_i>$ are the equivariant $\rho_i^{\otimes 1}$ density expansion coefficients we have considered above.

\begin{figure}
\centering
\includegraphics[width=1.0\linewidth]{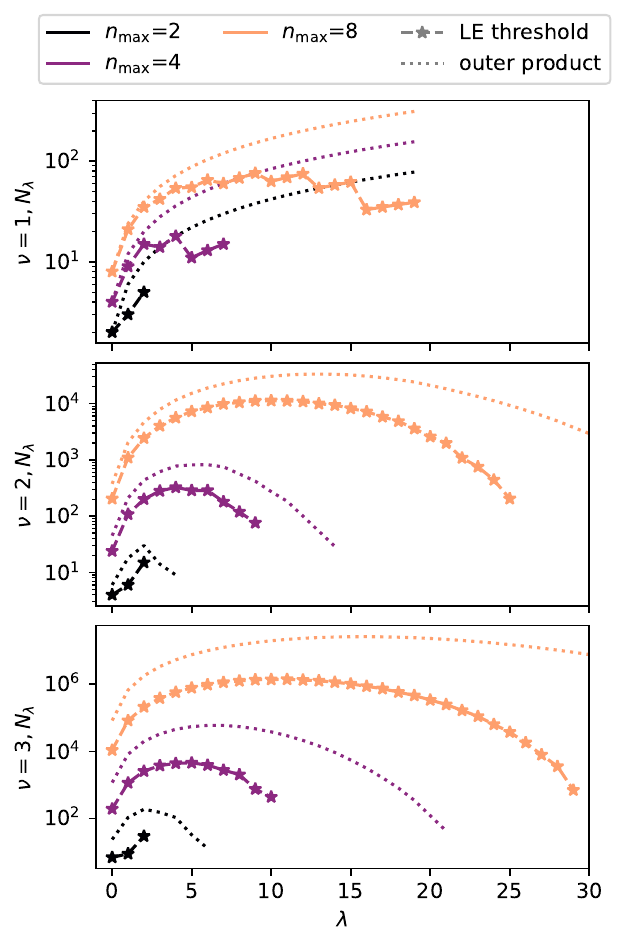}
\caption{Number of $\rep<q||\frho[\lambda\mu]_i^{\nu}>$ features retained for each equivariant channel $\lambda$, with (top to bottom) $\nu=1,2,3$. Dashed lines correspond to the features retained by applying the threshold $E_{\rm max}(\nu)=E_{\rm max}(1)+(\nu-1)E_{10}$ at each iteration, with $E_{\rm max}(1) = E_{n_{\rm max},0}$. Dotted lines correspond to taking the outer product density basis for $\nu=1$, and the outer product of the lower-order LE features for $\nu=2,3$. 
} 
\label{fig:le-sz-nu123}
\end{figure}

The product of these separable coefficients is an order-$\nu$ equivariant descriptor in the uncoupled angular momentum basis, 
\begin{equation}
 \prod_{k=1}^\nu c^i_{n_k l_k m_k} \equiv \rep<\nlm_1; \ldots \nlm_{\nu}||\rho_i^{\otimes\nu}>.
\label{eq:uncoupled}
\end{equation}
As discussed in Ref.~\citenum{niga+20jcp}, these can be transformed into the irreducible coupled form $\rep<q||\frho[\sigma;\lambda\mu]_i^\nu>$, which can also be constructed iteratively using Eq.~\eqref{eq:gen-cg-iter}. 
Considering the iterative construction, it is easy to see that at each order the coupling combines the projections of degenerate LE functions with the same values of $n$ and $l$, and so the irreducible SO(3) equivariant basis states are also eigenstates of the Laplacian in $\Omega^{\nu}$.

It follows that an ACDC construction based on LE functions will generate the smoothest possible \rev{(in the sense of a multi-dimensional generalization of Eq.~\eqref{eq:rq})} equivariant basis in which to expand the $\nu$-neighbor density correlations, provided the expansion is truncated appropriately. In particular, smoothness will be achieved at each order $\nu$ by retaining all equivariants with $\Omega^{\nu}$ Laplacian eigenvalues below some threshold $E_{\rm max}({\nu})$. The $\nu$-independent choice 
\begin{equation}
E_{\rm max}(\nu)=E_{\rm max}(1)+(\nu-1)E_{10}
\label{eq:emaxnu}
\end{equation}
is especially convenient because it ensures that every LE basis function with $E_{nl}\le E_{\rm max}(1)$ is included in at least one $\nu$-neighbor equivariant, and that no LE basis function with $E_{nl} > E_{\rm max}(1)$ is ever used.

Fig.~\ref{fig:le-sz-nu123} shows that the application of this threshold at each iteration reduces by at least an order of magnitude the number of equivariants relative to the number obtained by computing all possible combinations of the $\nu=1$ features with $E_{nl}\le E_{\rm max}(1)$. \rev{An additional reduction in the number of features can be achieved by applying selection rules that discard linearly dependent terms, such as considering only lexicographically-sorted $(n,l)$ terms\cite{niga+20jcp}. }
However, this still does not eliminate the exponential scaling of the number of features with $\nu$. An increasingly aggressive truncation might therefore be preferable for larger $\nu$, especially in situations where higher body-order terms are only expected to make a small contribution to the structure-property relations that are the goal of the calculation.

\subsubsection{Residual variance}\label{High-order-residual-variance}

The selection of high-$\nu$ equivariants on the basis of their 
$\Omega^{\nu}$ Laplacian eigenvalues is clearly only a viable option for the LE basis: it cannot be done for the metric-optimized X-OPT and J-OPT basis sets introduced in Sec.~II.D. There are, however, alternative ways to select the high-$\nu$ features for these basis sets,\cite{niga+20jcp,imba+18jcp,cers+21mlst} as we shall now briefly describe for the case of the X-OPT (minimal residual variance) basis.

The basic idea is to retain as much variance as possible in the high-$\nu$ density expansion coefficients, just as the X-OPT basis retains maximum variance in the $\nu=1$ coefficients. The limit on the amount of variance that can be retained with a finite basis is determined by the CG iteration in Eq.~\eqref{eq:gen-cg-iter}, which implies that the square modulus of each combination of features is the product of the square moduli of the lower-order features being combined:\cite{gosc+21jcp} 
\begin{multline}
\sum_{l_1l_2}\sum_{\sigma\lambda\mu} \left| \rep<ql_2; nl_1||\frho[\sigma;\lambda\mu]_i^{{\nu+1}}> \right|^2 =
\\
\sum_{l_1m_1} \left|\rep<n||\frho[1;l_1m_1]_i^1>\right|^2
 \sum_{\sigma_2l_2m_2}\left|\rep<q||\frho[\sigma_2;l_2m_2]_i^{\nu}>\right|^2.
\label{eq:norm-product}
\end{multline}
Clearly, therefore, the fraction of the variance that can be retained in the high-$\nu$ equivariants of a finite $nlm$ basis decreases exponentially with increasing $\nu$. The goal is nevertheless to retain as much of this limiting variance as possible, while at the same mitigating the exponential increase in the number of equivariants with $\nu$.

The $N$-body iterative contraction of equivariants (NICE) scheme\cite{niga+20jcp} does this by performing a principal component analysis (PCA), and then evaluating the combinations with the highest variance so as to yield the optimal features for a given size of the pool of ACDC equivariants. However, this {\em a posteriori} contraction scheme requires the evaluation of a large number of features before applying the projection step, which can become quite expensive. We have therefore also tested a cheaper iterative variance optimization (IVO) scheme that pre-selects a smaller number of high-order features and functions in a similar way to the iterative CUR selection of Refs.~\citenum{imba+18jcp,cers+21mlst}. This IVO scheme is described in Appendix~\ref{app:ivo} and its high-$\nu$ equivariants are compared with those of the PCA scheme for the X-OPT basis in our benchmark calculations in Sec.~III.

\subsubsection{Jacobian condition number}

Consider the Jacobian ${\bf J}_i$ computed for an equivariant of body order $\nu$, with elements $J_{q\sigma\lambda\mu,ij\alpha}=\rep<q||\field[\sigma; \lambda\mu]{\partial_{j\alpha} \rho}_i^\nu>$. As we show in Appendix~\ref{app:hi-jacobian}, in the sharp Gaussian smoothing and complete basis set limit, $\mathbf{J}_i^T\mathbf{J}_i$ can be evaluated in closed form, yielding 
\begin{equation}
\!\!\!\sum_{q\sigma\lambda\mu}
\rep<q||\field[\sigma;\lambda\mu]{\partial_{j\alpha}\rho}_i^\nu>
\rep<q||\field[\sigma;\lambda\mu]{\partial_{j'\alpha'}\rho}_i^\nu>=
k_i\delta_{jj'}\delta_{\alpha\alpha'},
\label{eq:highnu-jcn}
\end{equation}
which again implies a condition number of one. [Note that this expression involves summing over multiple equivariant channels, and that in general \emph{invariant} features cannot have uniform sensitivity: for high-symmetry structures (and some non-symmetric ones, in the case of degenerate low-order descriptors) $\rep<q||\field[1;00]{\partial_{j\alpha}\rho}_i^\nu>$ is strictly singular.\cite{pozd+21ore}] 

\section{Benchmarks}
\label{sec:benchmarks}

We shall now compare the performance of the LE basis with that of the data-driven X-OPT and J-OPT bases and the Gaussian-type orbital (GTO) basis presented in Ref.~\citenum{musi+21jcp}. To do this we shall use the three metrics introduced in Section \ref{sec:metrics}, as well as  regression exercises involving DFT potential energies as the supervised target.

\subsection{Chemical datasets and parameters}

We have performed benchmark calculations on the AIRSS \cite{pick-need11jpcm} carbon dataset \cite{pickard_2020_airss} and the random methane dataset\cite{pozdnyakov_2020_randomlydisplaced} (using carbon-centered environments only). The random selection of carbon-centered environments in the methane dataset provides a highly uniform distribution of neighbors, whereas the AIRSS carbon structures are more representative of the sort of systems for which potential energies are typically learned. A subset of the 10,000 lowest-energy structures in the random methane dataset was also employed to illustrate how a data-driven basis can be optimized for predominantly tetrahedral structures. A tetrahedral order parameter\cite{chau_1998_a} is used in the SI to establish the tetrahedral character of the low-energy methane structures. 

The cutoff radius $\rcut$ was set to $3.5$ \AA \ for the random and low-energy methane datasets, and to $5.0$ \AA \ for the condensed phase carbon dataset so as to include more coordination shells.
Gaussian-smoothed densities were used, with a Gaussian smoothing parameter of $\sigma=0.2$ \AA. Since all the C-H distances in the methane dataset are below $r_{\rm max}=3$ \AA, the hydrogen density with this smoothing parameter is essentially zero at $r_{\rm cut}=3.5$ \AA, so we used a radius of $a=r_{\rm cut}$ when defining the LE basis functions. For the learning exercises on the condensed phase carbon dataset, we also used $a=r_{\rm cut}$, but introduced a cutoff function to damp the density to zero at this radius. The precise form of the cutoff function is given along with the other details of the LE expansion in Appendix~A.  

\begin{figure}[b]
\centering
\includegraphics[width=\linewidth]{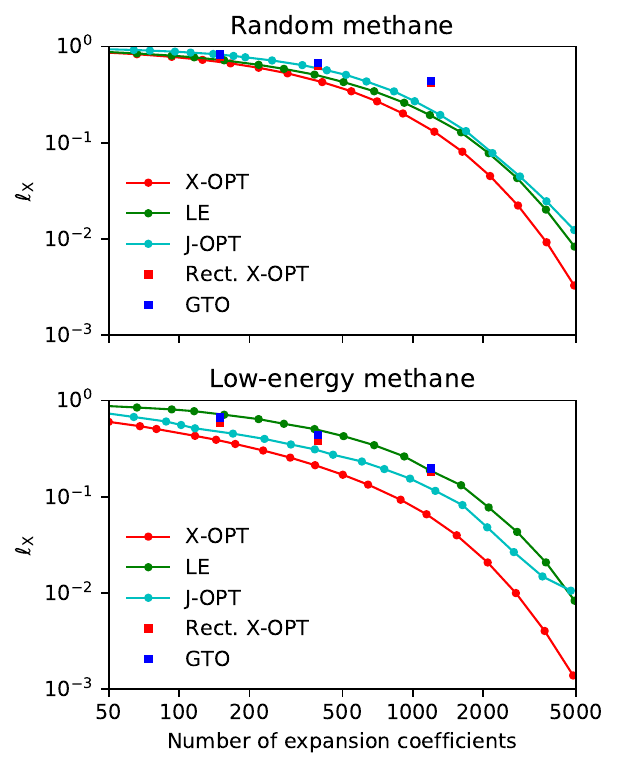}
\caption{Residual variances of different bases on methane structures. \rev{$\ell_X$ is the normalised $L^2$ error in the approximation to the Gaussian-smoothed density in Eq.~\ref{eq:variance}. The curves are the results obtained with the X-OPT, LE, and J-OPT basis sets truncated to the specified number of $(n,l)$ basis functions (or expansion coefficients) as described in the text. The square markers correspond to full outer-product X-OPT and GTO basis sets with $(n_{\text{max}},l_{\text{max}})=(6,4), (8,6)$ and (12,9). } }
\label{fig:carbon-variance}
\end{figure}

The LE expansion was truncated by retaining basis functions with $E_{nl}\le E_{\rm max}$ as described in Sec.~II.B. The more traditional GTO and ``rectangular'' X-OPT bases, which correspond to using rectangular cutoffs in Fig.~\ref{fig:nmaxl}, were truncated via three representative $(n_{\rm max},l_{\rm max})$ combinations (small $(6,4)$, medium $(8,6)$, and large $(12,9)$), and they are indicated by square markers in our figures.

In what follows we report only selected results that highlight the most important observations for each type of benchmark. 
A more comprehensive series of plots, together with a more detailed discussion, can be found in the SI.

\subsection{Density expansion coefficients}

\begin{figure}[b]
\centering
\includegraphics[width=\linewidth]{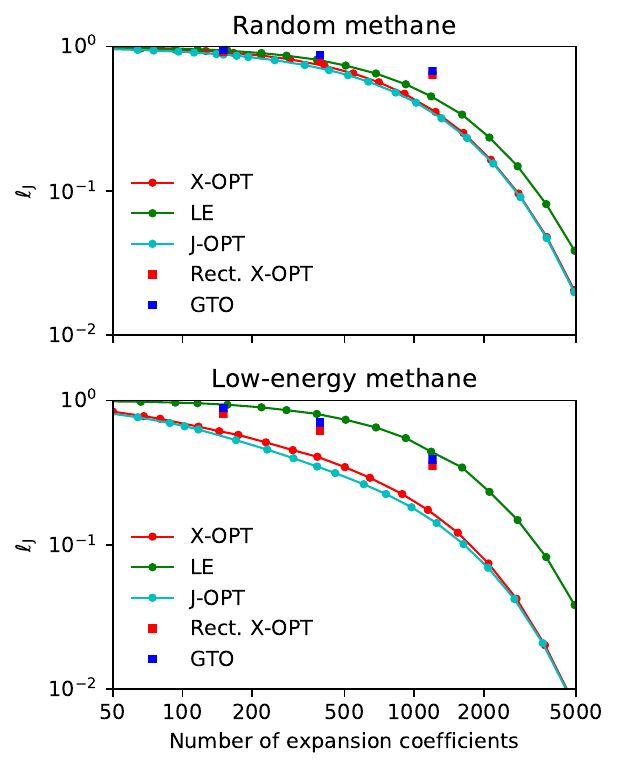}
\caption{Residual Jacobian variances of different bases on methane structures. \rev{$\ell_J$ is the error in approximation to the Jacobian of the density expansion defined in Eq.~\ref{eq:Jacobianvariance}. The basis sets are the same as those in Fig.~\ref{fig:carbon-variance}.}}
\label{fig:carbon-jac-variance}
\end{figure}

\subsubsection{Residual variance}

The residual variance tests conducted on methane are shown in Fig. \ref{fig:carbon-variance}. The X-OPT basis (unsurprisingly) performs best for this metric, while the rectangular bases generally suffer from the exclusion of high-$l$ channels.
The LE basis is competitive with all the others, and, given an X-OPT basis of a certain size, it only takes a slightly larger LE basis to match it in terms of captured variance. The benefits of using a data-driven basis are more pronounced for the low-energy methane configurations, in which the \ce{H} atoms are non-uniformly distributed around the central carbon.

\subsubsection{Jacobian variance}

The Jacobian variance tests for methane shown in Fig. \ref{fig:carbon-jac-variance} confirm that the J-OPT basis is optimum for this alternative metric. Once again, the $(n_{\rm max}, l_{\rm max})$-truncated bases suffer from the lack of high-$l$ coefficients. The LE basis is competitive with the data-adapted bases for random methane, although noticeably less so for the heavily optimizable low-energy methane structures.

Comparing  Figs.~\ref{fig:carbon-variance} and~\ref{fig:carbon-jac-variance} reveals the trade-off between optimizing the variance and the Jacobian variance: although one has to choose one target property to optimize, the resulting basis set also performs well for the other task. This is especially true of the X-OPT basis, which does remarkably well on the Jacobian variance test. Another key observation is that tuning the number of radial functions within each $l$ channel results in a systematically more efficient encoding of information than is present in an outer-product basis of comparable size. This is particularly evident when comparing the two truncation methods for the X-OPT basis. When using the same number of radial functions for each angular momentum channel, the improvement over a primitve GTO basis is barely noticeable, whereas using an adaptive truncation of the radial basis yields an improvement of a factor between 2 and 5 in both $\ell_X$ and $\ell_J$.

\begin{figure}[b]
\centering
\includegraphics[width=\linewidth]{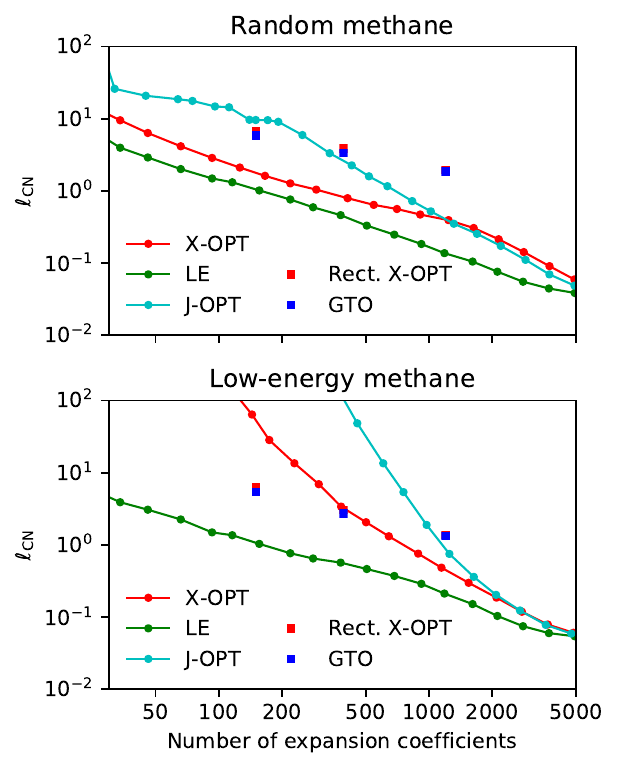}
\caption{Jacobian condition number tests on methane structures. \rev{$\ell_\text{CN} \rightarrow 0$ indicates that the condition number of the density-expansion Jacobian tends to one. The basis sets are the same as those in Fig.~\ref{fig:carbon-variance}.}}
\label{fig:carbon-cn}
\end{figure}

\subsubsection{Jacobian condition number}

The Jacobian condition number (CN) tests are shown in Fig.~\ref{fig:carbon-cn}. In contrast to the variance and Jacobian variance tests, these CN tests show the LE basis outperforming all of the other basis sets we have considered. Because it is smooth in the entire atomic neighborhood,  even a small LE basis yields a uniform approximation to the atom density, and therefore small anisotropy in the sensitivity of the features to structural deformations, even though it provides a worse $L^2$ approximation to the Jacobian (see Fig.~\ref{fig:carbon-jac-variance}). The X-OPT basis has a comparable CN to the LE basis for the random methane data set, but a much higher CN for the more optimizable low-energy methane subset. This behaviour can be understood by noting that adapting the basis to the data set makes it less uniform, leading to a more anisotropic response to deformations. A similar reasoning explains why the J-OPT basis, which provides a better approximation to the density derivatives, yields an even higher CN (and therefore a worse result still in this respect) than the X-OPT basis.

\subsection{Equivariant \texorpdfstring{$\nu=2$}{Lg} features}

The density expansion coefficients ($\nu=1$ equivariant features) are the starting point of all density-correlation descriptors. In order to build accurate ML models, however, it is almost mandatory to increase the correlation order beyond $\nu=1$.
To investigate the impact of the body-order iteration on the different metrics we have introduced, we have repeated the  residual variance and Jacobian condition number tests using $\nu=2$ equivariants. 
For this we have confined our attention to the LE and X-OPT bases, because they perform well on the other tests and their associated truncation criteria readily generalize to higher body-order features.

\begin{figure}[b]
\centering
\includegraphics[width=\linewidth]{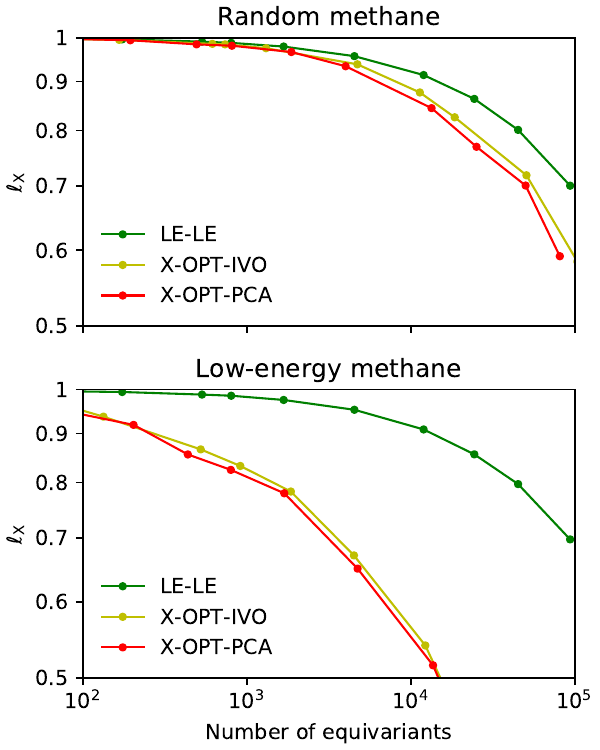}
\caption{Residual variances of $\nu=2$ equivariants on methane structures. \rev{ The LE-LE data corresponds to density correlations built on the LE basis, and truncated based on the LE eigenvalues. The X-OPT-PCA and X-OPT-IVO data correspond to a variance-optimal density expansion, with correlations further contracted using PCA or selected using the IVO scheme. }  }
\label{fig:carbon-nu2-variance}
\end{figure}

\subsubsection{Residual Variance}

The residual variances of the LE and X-OPT $\nu=2$ equivariant features for the random and low-energy methane structures are shown in Fig.~\ref{fig:carbon-nu2-variance}. The PCA contraction retains more variance than the LE truncation, as is obvious from its construction. Contrary to the $\nu=1$ case, however, the retained variance gap is large, especially for the low-energy methane structures, indicating that the $\nu=2$ equivariants are significantly more compressible than the $\nu=1$ density expansion coefficients. 

From a numerical point of view, this comparison is not entirely fair, because while the X-OPT basis can be evaluated at no additional cost\cite{gosc+21jcp} compared to the LE basis, the PCA contraction of higher-order representations requires the evaluation of all features before contracting them. This makes it more expensive than using Eq.~(29) to pre-select the relevant higher-order equivariant features for the LE basis, as we have done in computing the LE-LE curves in Fig.~\ref{fig:carbon-nu2-variance}.
The IVO scheme, which is based on the selection of the highest variance features, is more closely comparable: after having identified the optimal components, it only requires evaluating those that have been chosen. The resulting IVO equivariants in Fig.~\ref{fig:carbon-nu2-variance} have slightly higher residual variance than those from PCA, but they are still significantly lower than those from LE-LE.

\begin{figure}[b]
\centering
\includegraphics[width=\linewidth]{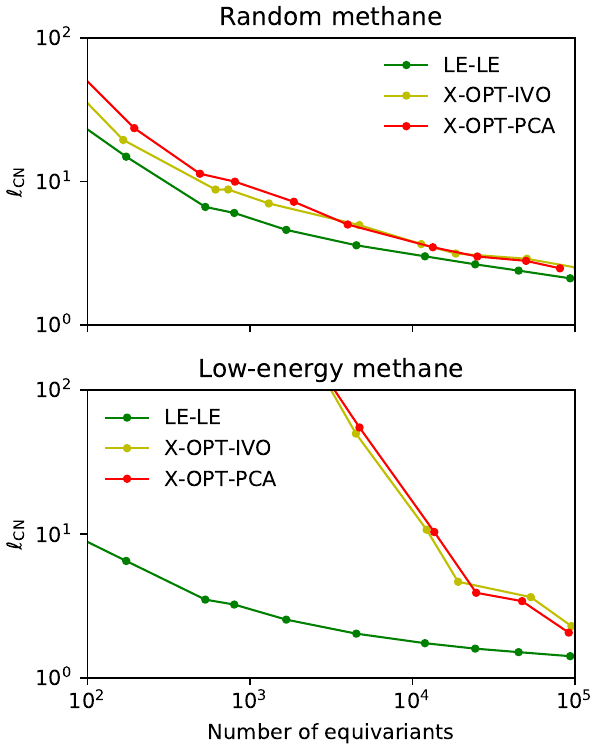}
\caption{$\nu = 2$ Jacobian condition number tests on methane structures. \rev{The LE-LE, X-OPT-PCA, and X-OPT-IVO expansions of the $\nu=2$ equivariants are the same as in the residual variance tests in Fig.~\ref{fig:carbon-nu2-variance}.} }
\label{fig:carbon-nu2-cn}
\end{figure}

\subsubsection{Jacobian condition number}

The Jacobian condition numbers of the $\nu=2$ equivariants from the X-OPT and LE methods are shown in Fig.~\ref{fig:carbon-nu2-cn}. As in the $\nu = 1$ case, the LE equivariants give the smallest CN, for both the random and the low-energy methane structures. The CN of the X-OPT equivariants increases significantly on going from the random to the low-energy structures, regardless of the feature reduction method employed (PCA contraction or IVO selection). This reinforces the notion that adapting the basis to focus on the correlations that are most common in a given data set increases -- rather than decreases -- the anisotropy of the Jacobian, at least in the small basis set limit. As more $\nu=2$ equivariants are included in the X-OPT calculation, the Jacobian CN decreases, and by the time there are 10$^5$ it appears to be approaching that of the LE calculation.

\begin{figure*}[t]
\includegraphics[width=0.995\textwidth]{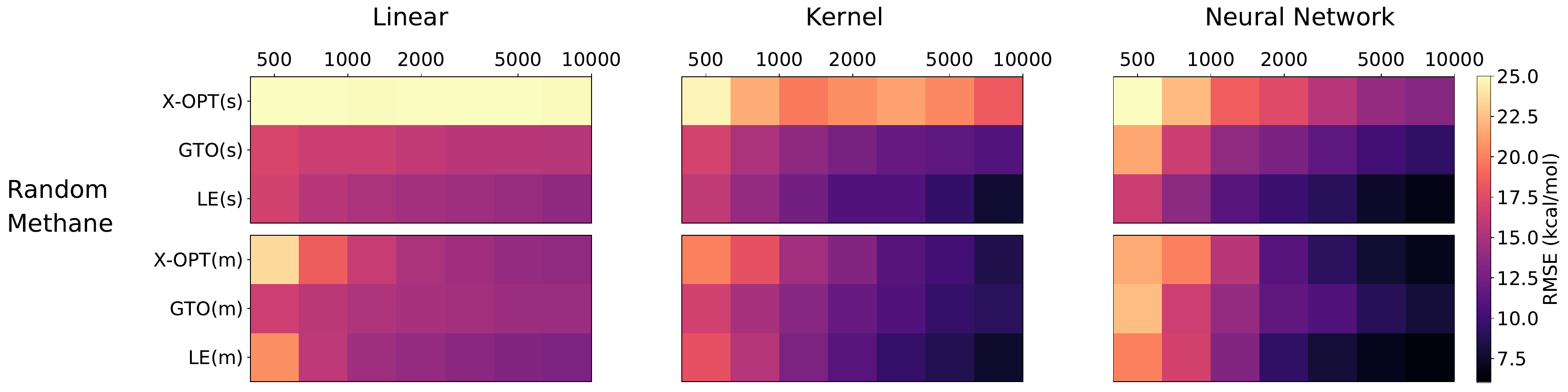}
\vspace{0.13cm}
\hspace*{-0.1cm}\includegraphics[width=0.985\textwidth]{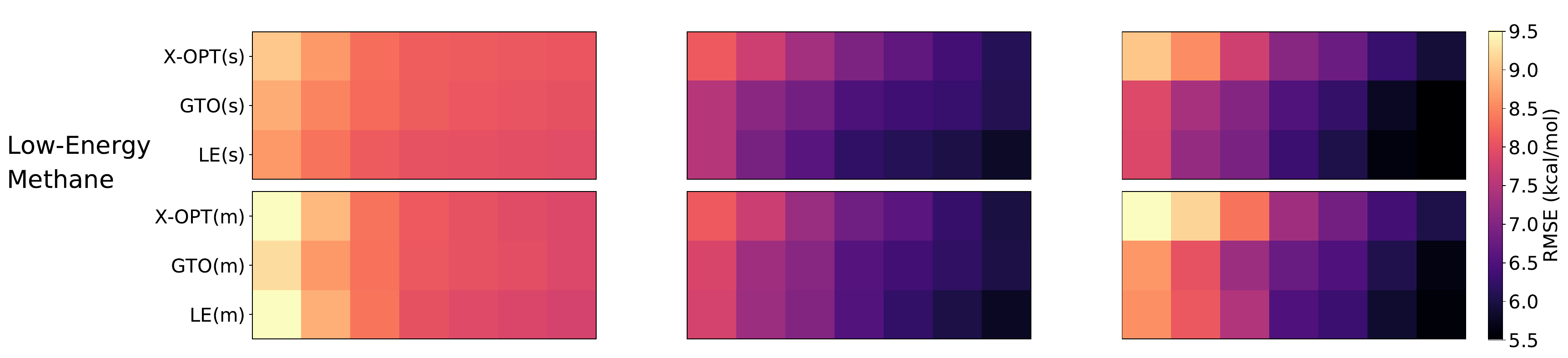}
\includegraphics[width=0.995\textwidth]{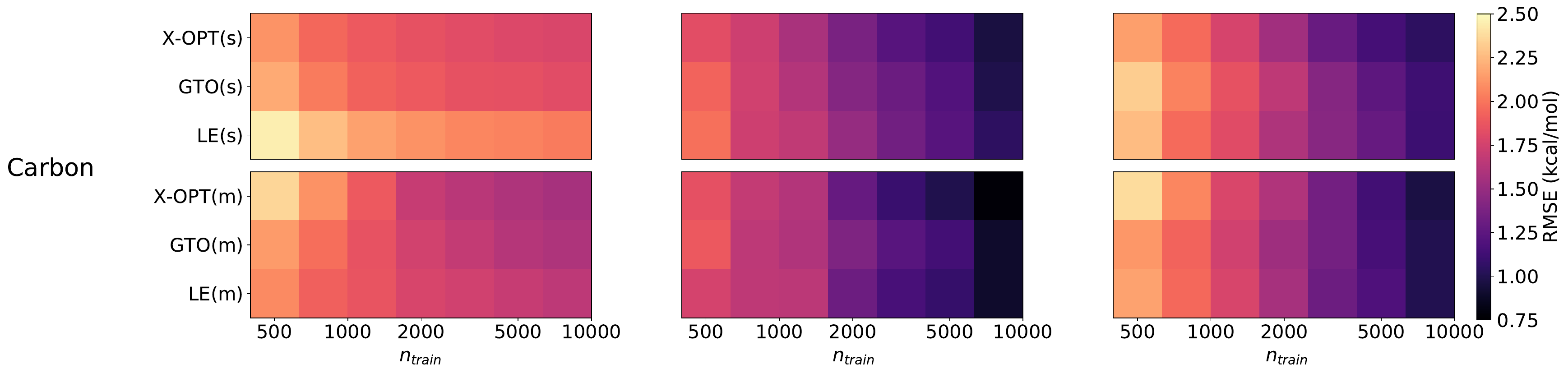}
\caption{Learning using invariants with $\nu \le 2$. (s) denotes a small basis, and (m) denotes a medium-sized basis. Each of these is constructed so as to contain a similar number of basis functions for all three methods (X-OPT, GTO, and LE), as described in the SI. Larger expansion coefficient sets exhibit significant overfitting in the case of methane (for these relatively small training set sizes), and they do not significantly improve on their medium-sized counterparts in the case of carbon (see the SI). Additional results for the J-OPT and rectangular X-OPT bases are also available in the SI.}
\label{fig:nu2learning}
\end{figure*}

\subsection{Learning from invariant features}

While the metrics discussed above provide an ``unsupervised'' estimate of the quality of the basis set used for the density expansion, the ultimate goal of most ML models is to achieve accurate predictions for a target property. 
We take as a paradigmatic example the construction of models of the interatomic potential, using only energies for training, comparing the performance of the different radial basis sets we have considered. As in the previous section, we show results for some representative cases, and refer the reader to the SI for further examples and a more comprehensive discussion. 

\subsubsection{Invariants with \texorpdfstring{$\nu\le 2$}{Lg}}\label{learning_nu2}

Fig.~\ref{fig:nu2learning} compares the performance of (up to) $\nu=2$ invariants built from different bases as inputs to linear, kernel, and neural network (NN) models. Linear-based learning saturates quickly due to its inability to account for high body-order interactions, while kernel and NN-based models do not suffer from this issue. 
The relative performance of different bases is similar across different machine learning models, highlighting a high degree of ``orthogonality'' between the choice of basis and that of the fitting model, at least for the datasets we have considered. 

There is comparatively little to choose between the X-OPT, GTO, and LE results in Fig.~\ref{fig:nu2learning} for either methane dataset -- random methane or low-energy methane -- regardless of whether the potential is learned from a linear fit, with a non-linear kernel, or with a neural network. The only real outlier is the small X-OPT basis, which performs particularly poorly for the random methane dataset. The \lq\lq optimization" of this basis to fit the density more accurately than a LE basis of comparable size clearly does not help with this learning exercise. Indeed, while the differences are often small, the LE results are slightly better than those of the other two basis sets in all 12 of the methane panels in Fig.~\ref{fig:nu2learning}.

The results of the carbon tests in Fig.~~\ref{fig:nu2learning} are another matter. This is a simpler problem that can be solved to \lq\lq chemical accuracy" (a root mean square error of less than 1 kcal/mol) using any of the three basis sets, provided either a non-linear kernel or a neural network is used to extract higher body-order contributions from the invariant features with $\nu\le 2$. Once this is done, the results for all three basis sets are essentially the same, provided an appropriate cutoff function is applied to the density to make it go smoothly to zero at $r_{\rm cut}$. [As we have done in these tests: for the LE basis, we used the cutoff functions $f_1(x)$ and $f_2(r_{ij})$ defined in Appendix~A. These were also used when constructing the primitive LE basis that was contracted to produce the X-OPT basis functions. For the GTO basis, $f_1(x)$ is not needed, so we simply used $f_2(r_{ij})$.]

The use of an aggressive cutoff function that emphasizes the short-range interactions, that are dominant for this problem, is essential to achieve the good performance seen in Fig.~\ref{fig:nu2learning}. As has been noted previously\cite{huan-vonl16jcp,will+18pccp,drau19prb}, this type of feature engineering, which incorporates prior knowledge on the target function, is very effective in improving the performance of a ML model. On the other hand, it appears that in this case the choice of basis set plays only a very limited role. The most accurate results for the carbon dataset in Fig.~\ref{fig:nu2learning} are seen for the X-OPT basis, which is optimized to fit the radially-weighted density once the cutoff function has been introduced. However, these results are only marginally better than those obtained with the LE basis, which is not optimized to fit the dataset at all.

\begin{figure}[hbpt]
\centering
\includegraphics[width=0.9\linewidth]{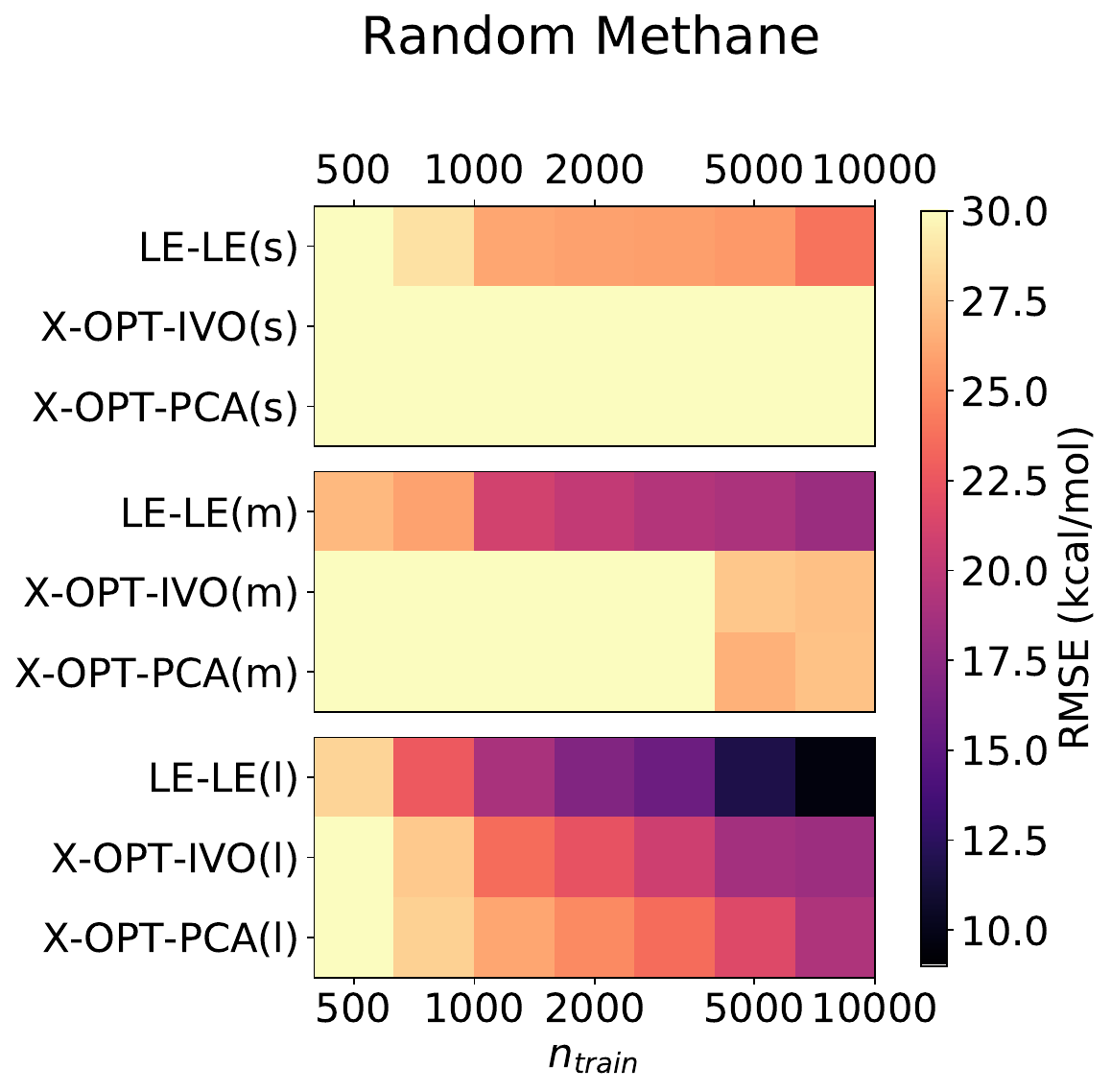}

\vspace{5mm}  %

\includegraphics[width=0.88\linewidth]{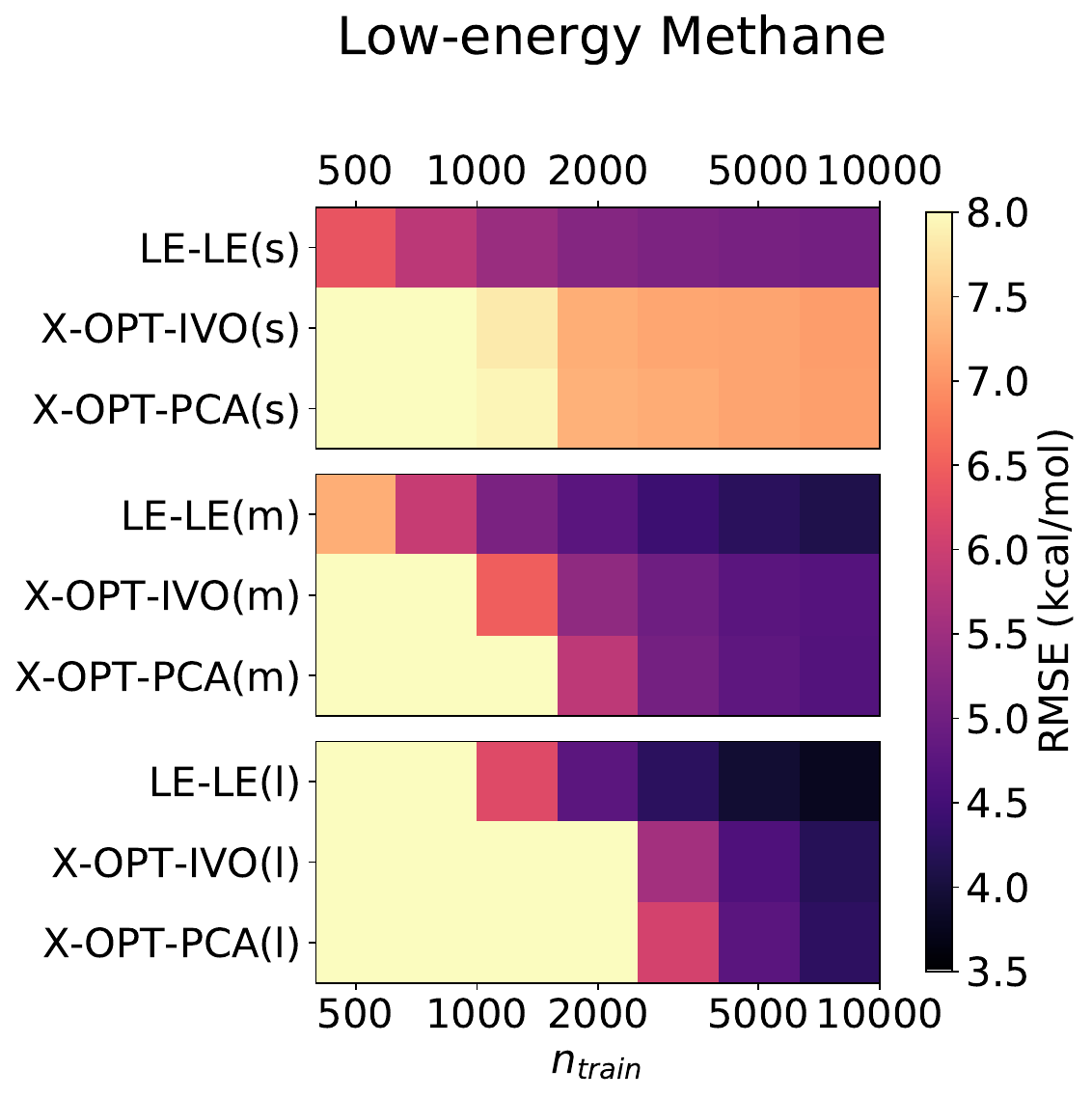}
\caption{Learning from up to $\nu = 3$ invariants. LE-LE denotes a LE basis combined with a Laplacian eigenvalue selection at the $\nu = 2$ equivariant level. Similarly, X-OPT-IVO indicates an X-OPT basis ($\nu = 1$) and an IVO contraction ($\nu = 2$), and X-OPT-PCA indicates PCA contractions at both $\nu = 1$ and $\nu = 2$. (s), (m), and (l) denote small-sized, medium-sized, and large $\nu = 3$ invariant sets, respectively.}
\label{fig:nu3learning}
\end{figure}

\subsubsection{Invariants with \texorpdfstring{$\nu\le 3$.}{Lg}}

Given the saturation of the linear learning curves in Fig.~\ref{fig:nu2learning}, the linear learning exercises for the methane datasets were repeated using (up to) $\nu = 3$ invariants. The results of these tests are shown in Fig.~\ref{fig:nu3learning}, where it is clear that, when enough $\nu = 3$ invariants are included, the curves do not saturate within the range of training set sizes considered. Note also that the use of the LE basis, together with the LE selection scheme for high-order equivariants (LE-LE), affords the lowest RMSEs in these tests for both methane datasets, by a substantial margin. 
This suggests that the optimum smoothness of the LE basis, which extends in a natural way to the smoothness of the higher body-order features obtained from the LE-LE scheme as described in Sec.~II.E, is an important ingredient to successfully learn potential energies from high-$\nu$ invariants. Since this seems to be the direction of travel of much of the most recent work on machine learning potentials,\cite{shap16mms, drau19prb, niga+20jcp} it could well be the most significant result in this paper, especially since the  data-driven and ostensibly \lq\lq optimized" X-OPT-IVO and X-OPT-PCA results in Fig.~\ref{fig:nu3learning} are so poor by comparison.

It is important to keep in mind here, though, that the methane dataset, particularly when learning with only C-centered descriptors, is particularly challenging, and heavily dependent on the convergence and resolution of the basis set. \rev{Even though the} advantage of the LE-LE scheme for other regression tasks, such as learning with both C- and H-centered descriptors, \rev{are} not be quite so pronounced, \rev{they are nonetheless significant (see the next Section, and especially Fig. \ref{fig:methane})}. \rev{In particular}, it is remarkable that a smoothness criterion, that leads to a very high loss of information content in terms of $\ell_X$ (see Fig.~\ref{fig:carbon-nu2-variance}), proves so effective when it comes to regression performance. This suggests that a complete description of the $\nu$-order density correlations might not be necessary to converge ML models based on them, and that actually seeking the most complete description possible (as is done in X-OPT-PCA) might not be the best approach after all.

\subsubsection{Application to body-ordered expansions}\label{sec:LE-ACE}

\rev{
The LE-LE idea presented in the previous section is easily extensible to body-ordered expansion frameworks\cite{drau19prb, niga+20jcp}. We will now show how the LE basis improves over previous basis functions when fitting interatomic potentials for the molecules in the rMD17 dataset.\cite{rMD17} Given that we will aim to compare with existing results based on the ACE scheme,\cite{kov2021jctc} we will incorporate some of the heuristics that have been used in previous ACE work. Thus, we refer to the resulting model as LE-ACE.

\begin{figure}[btp]
\centering
\includegraphics[width=1.0\linewidth]{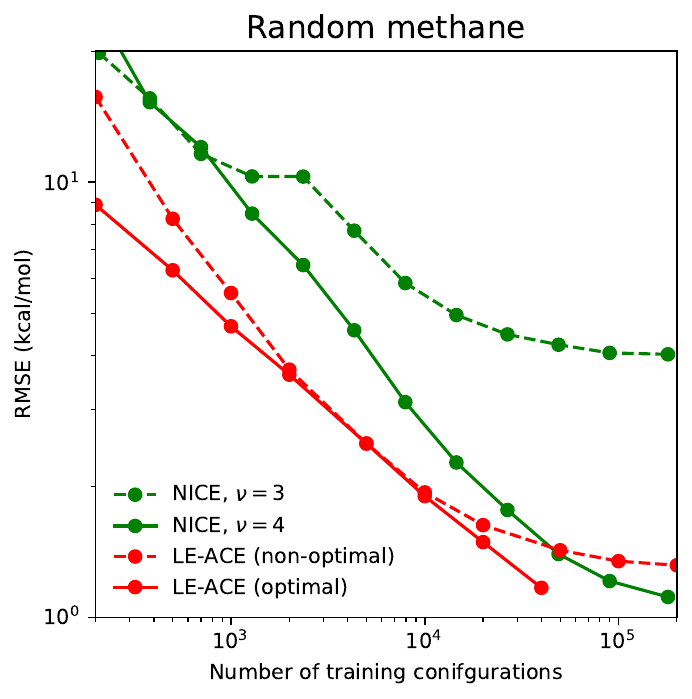}
\caption{Performance of the LE-ACE model on the random methane dataset. The NICE curves are from Ref.~\citenum{niga+20jcp}. The optimal truncation of LE-ACE follows the $E_\textrm{max}{(\nu)} \propto n_\textrm{train}^{2/3\nu}$ scaling, while the non-optimal truncation uses the optimal values $E_{\rm max}(\nu)$ for 5000 training structures and extends them to all other points on the learning curve.
} 
\label{fig:methane}
\end{figure}

Before proceeding to this comparison, we note that one of the challenges when benchmarking linear models is that validation errors depend strongly on regularization, and in particular they depend in a non-monotonic way on the size of the feature vector. For example, when using a small training set one often observes that aggressive truncation results in more transferable models.
In the LE-LE context, this means that the $E_\textrm{max}{(\nu)}$ values should be optimized as a function of $n_\textrm{train}$. This follows because the maximum frequency component of the learned potential at each body order depends on $E_{\rm max}(\nu)$. As more samples are included in the training set, the configuration space of the system is sampled more densely, and therefore higher-frequency components of the potential can be learned. Including basis functions whose frequency is too high leads to overfitting, whereas if $E_\textrm{max}{(\nu)}$ is smaller than optimal there will be underfitting. 
   
In order to estimate how $E_\textrm{max}(\nu)$ should vary as a function of $n_\textrm{train}$, we note that $E_\textrm{max}{(\nu)}$ is proportional to the square of the maximum sampled frequency $\omega$ in the $\Omega^\nu$ hypersphere. Since $\omega$ is proportional to the inverse of the typical distance $d$ between configurations in the $\Omega^\nu$ space, we have $E_\textrm{max}{(\nu)} \propto \omega^2 \propto 1/d^2$. If we consider one training point to represent one point in the $\Omega^\nu$ hypersphere, the typical distance $d$ scales as $(V^{\nu}/n_\mathrm{train})^{1/3\nu}$, where $V^{\nu}$ is the volume of $\Omega^\nu$. Hence, $E_\textrm{max}{(\nu)} \propto n_\mathrm{train}^{2/3\nu}$.

We can illustrate this idea for the random methane dataset by extending the training set beyond the 10,000 configurations we have used in most of the present work. 
As shown in Fig.~\ref{fig:methane}, the proposed scaling of $E_\textrm{max}{(\nu)}$ gives rise to a learning curve that decays linearly on a log-log scale. In contrast, all other learning curves are non-optimal. The $\nu = 3$ NICE curve\cite{niga+20jcp} shows limited accuracy and saturation due to its inability to describe 5-body interactions in the gas-phase methane molecules. The $\nu = 4$ NICE and non-optimal LE-ACE curves show signs of saturation (underfitting) in the data-rich regime, as well as overfitting in the data-poor regime, due to their sub-optimal resolution. By contrast, the newly proposed $E_\textrm{max}{(\nu)} \propto n_\mathrm{train}^{2/3\nu}$ truncation provides a good fit regardless of the amount of training data, suggesting that our frequency-based argument provides a valid heuristic to adapt the resolution of a linear model to the amount of data available.

We will now compare the performance of this new model with that of a standard ACE implementation on the rMD17 dataset\cite{rMD17}, which is widely used as a benchmark, and allows us to compare LE-ACE with several recently-proposed models. 
This is however a somewhat problematic data set, as it contains highly-correlated structures generated from short, low-temperature molecular dynamics trajectories: as noted in Ref.~\citenum{mace}, the most instructive comparisons can be obtained in the data-poor regime.
Further details regarding our implementation and additional benchmarks are provided in Appendix \ref{app:le-ace} and Section 4 of the SI. We should emphasize here that, for fairness of comparison, we have incorporated a heuristic optimization in the form of a non-linear transformation of the radial coordinate [see Eq.~\ref{eq:radialtransform}], as is almost universally done when learning interactomic potentials in ACE \cite{drau19prb, kov2021jctc, dusson_2022_atomic}.

\begin{table}[bthp]
\centering
\begin{tabular}{lccccc} 
\hline\hline
Molecule & Target & ACE & LE-ACE & NequIP & MACE  \\
\hline
\multirow{2}{*}{Aspirin}        & E & 26.2 & 22.4(2.0) & 19.5 & \textbf{17.0} \\
                                & F & 63.8 & 59.1(1.3) & 52.0 & \textbf{43.9} \\
\hline
\multirow{2}{*}{Azobenzene}     & E & 9.0  & 9.9(8) & 6.0 & \textbf{5.4}  \\
                                & F & 28.8 & 27.5(6) & 20.0 & \textbf{17.7} \\
\hline
\multirow{2}{*}{Benzene}        & E & 0.2  & \textbf{0.135(9)} & 0.6 & 0.7 \\
                                & F & 2.7  & \textbf{1.44(3)} & 2.9 & 2.7 \\
\hline
\multirow{2}{*}{Ethanol}        & E & 8.6  & \textbf{6.6(6)}  & 8.7 & 6.7  \\
                                & F & 43.0 & \textbf{32.0(1.5)} & 40.2 & 32.6 \\
\hline
\multirow{2}{*}{Malonaldehyde}  & E & 12.8 & 11.3(1.9) & 12.7 & \textbf{10.0} \\
                                & F & 63.5 & 50.9(3.6) & 52.5 & \textbf{43.3} \\
\hline
\multirow{2}{*}{Naphthalene}    & E & 3.8  & 2.9(1)  & \textbf{2.1} & \textbf{2.1}  \\
                                & F & 19.7 & 13.9(3) & 10.0 & \textbf{9.2}  \\
\hline
\multirow{2}{*}{Paracetamol}    & E & 13.6 & 14.3(1.8) & 14.3 & \textbf{9.7}  \\
                                & F & 45.7 & 45.1(1.6) & 39.7 & \textbf{31.5} \\
\hline
\multirow{2}{*}{Salicylic acid} & E & 8.9  & 8.3(4) & 8.0 & \textbf{6.5}  \\
                                & F & 41.7 & 36.7(1.4) & 35.0 & \textbf{28.4} \\
\hline
\multirow{2}{*}{Toluene}        & E & 5.3  & 4.1(3)  & 3.3 & \textbf{3.1}  \\
                                & F & 27.1 & 18.4(4) & 15.1 & \textbf{12.1} \\
\hline
\multirow{2}{*}{Uracil}         & E & 6.5  & 5.7(4) & 7.3 & \textbf{4.4}  \\
                                & F & 36.2 & 30.7(1.3) & 40.1 & \textbf{25.9} \\
\hline\hline
\end{tabular}   
\caption{Benchmark results on the rMD17 dataset\cite{rMD17}, in the smaller and more challenging version presented in Ref. \citenum{mace}, where the training set consists of only 50 structures. The table contains the mean absolute errors of energies (E) and force components (F) in units of meV and meV/\AA, respectively, as obtained using ACE [Ref.~\citenum{kov2021jctc}], LE-ACE [Present Work], NequIP [Ref.~\citenum{batzner2021se3}] and MACE [Ref.~\citenum{mace}]. The numbers in parentheses after the LE-ACE results are statistical errors in the final digit(s), which represent the standard deviation of five different train/test splits (see SI). The best results for each target are marked in bold.}
\label{tab:LE-ACE}
\end{table}

Our LE-ACE results for the rMD17 dataset are shown in Table~I, where they are compared with the results of a standard ACE implementation\cite{kov2021jctc} and those of two state-of-the-art equivariant message-passing neural networks (NequIP\cite{batzner2021se3} and MACE\cite{mace}). While there might be other minor differences between the two implementations, it can be seen that the use of the LE basis allows for a significant and systematic improvement in accuracy over the basis functions used in ACE. In some cases, the LE-ACE model is also competitive with NequIP and MACE. It is reasonable to assume that the incorporation of the ideas that underlie the LE basis into NequIP and MACE would lead to similar improvements in the message-passing context to the improvements LE-ACE offers over ACE.
}

\section{Conclusions}

In this paper we have investigated the use of the LE basis for the expansion of local atomic densities, explained why this basis, when truncated with the eigenvalue criterion $E_{nl}\le E_{\rm max}$, provides the smoothest possible basis of a given size within a sphere $\Omega$ \rev{in the sense of minimising the Rayleigh quotient in Eq.~\ref{eq:rq}}, and described how this property extends in a natural way to the representation of density correlations of arbitrary order $\nu$ within the relevant hyperspheres $\Omega^{\nu}$. 
Several of these developments have precedents in the recent atomistic machine learning literature. For example, the truncation criterion $E_{nl}\le E_{\rm max}$ leads to the retention of different numbers of radial basis functions in each angular momentum channel $l$, as is increasingly done when using atomic cluster expansion (ACE) descriptors.\cite{doi:10.1021/acs.jctc.1c00647, dusson_2022_atomic} 
The key difference is that the $E_{nl}\le E_{\rm max}$ criterion tells us exactly how many radial basis functions to retain for each value of $l$, without the need to introduce any further parameters. \rev{We have also described how an analogous criterion can be used to truncate higher body-order features of the LE basis [see Eq.~\ref{eq:emaxnu}], and provided a scaling argument that shows how to adapt the parameters in this more general truncation scheme to the training set size [$E_{\rm max}(\nu)\propto n_{\rm train}^{2/3\nu}$ -- see Sec.~III.D.3].}

We began by arguing that a smooth basis is desirable for machine learning because it implies a smooth interpolation (and possibly extrapolation) of machine-learned properties to configurations that are not present in the training set. Our learning results in Figs.~\ref{fig:nu2learning}-- \ref{fig:methane} certainly support this argument.
In all cases, the LE basis gives results are either comparable to or better than those obtained with data-driven basis sets obtained by optimizing the $L^2$ fit to the density, which we find results in a less uniform sensitivity of the density coefficients to atomic displacements.

The advantage of the LE basis is especially apparent in the learning results in Fig.~\ref{fig:nu3learning}, in which the $\nu\le 3$ invariants are built from $\nu\le 2$ equivariants that benefit from the smoothness of the LE representation of the density correlations in $\Omega^2$. The fact that this advantage is obtained when the LE $\nu=2$ equivariants only capture a small fraction of the variance in the $\nu=2$ density correlations present in the dataset (Fig.~\ref{fig:carbon-nu2-variance}) is especially interesting, as it suggests that a high resolution representation of the density correlations may not be needed to learn the potential energy. 

\rev{Our learning results are anti-correlated with the results of our residual variance ($\ell_X$) tests in Figs.~\ref{fig:carbon-variance} and~\ref{fig:carbon-nu2-variance}, but well correlated with the results of our Jacobian condition number ($\ell_{CN}$) tests in Figs.~\ref{fig:carbon-cn} and \ref{fig:carbon-nu2-cn}. While this does not imply that the Jacobian condition number is the ultimate indicator of basis set quality, we do believe that it points at the importance of achieving a discretization of the density that has ``uniform resolution'', and thereby does not artificially distort the sensitivity of the features to atomic deformations. 

A ``signal-processing'' view, in which features need to provide a basis that fully describes the density correlations, implies an exponential increase in the number of features with body order $\nu$.
If instead it suffices to retain enough features to guarantee that the Jacobian is full rank and well-conditioned for every configuration, then one may hope that a number of features proportional to the maximum number of neighbours within $r_\text{cut}$ would suffice. 
Our learning results in Figs.~\ref{fig:nu2learning}--\ref{fig:methane} suggest that, when it comes to building interatomic potentials, a basis and a truncation strategy that are optimal in a signal-processing sense are less effective than a basis and a truncation strategy aiming for \lq\lq optimal smoothness", which is the approach we have taken in this paper. 
}

\subsection*{SUPPLEMENTAL INFORMATION}
See supplemental information for a more comprehensive set of benchmarks, its analysis, and a detailed account of the Rayleigh-quotient formulation of optimized bases.

\subsection*{AUTHOR DECLARATIONS}
The authors have no conflicts of interest to disclose. 

\begin{acknowledgments}
\rev{The authors thank Sergey Pozdnyakov for sharing benchmarks of the NICE model on the random methane dataset.}
Michele Ceriotti and Kevin Huguenin-Dumittan acknowledge funding from the European Research Council (ERC) under the European Union’s Horizon 2020 research and innovation programme (grant agreement No 101001890-FIAMMA).
\end{acknowledgments}

\subsection*{AUTHOR CONTRIBUTIONS}
{David Manolopoulos} and {Michele Ceriotti} conceived and supervised this project. {Filippo Bigi} developed the LE smoothness argument and performed and analyzed the calculations. {Kevin Huguenin-Dumittan} developed the Rayleigh quotient view of the basis optimization problem. All authors contributed to the analytical derivations and to the writing of the paper.

\subsection*{DATA AVAILABILITY}
The data that support the findings of this study are available in the paper and the supplementary material. The datasets we use have been published elsewhere and are available from Refs.~\citenum{matcloud20a,matcloud20c,rmd17data}. 

\providecommand{\noopsort}[1]{}

\appendix

\section{Details of the LE basis construction}
\label{app:le-details}
The radial basis functions of the LE basis in a sphere of radius $a$ are
\begin{equation}
R_{nl}(x) = a^{-3/2}N_{nl}j_l(z_{nl}x/a),
\end{equation}
where $j_l(x)$ is a spherical Bessel function, $z_{nl}$ is its $n$-th zero, and
\begin{equation}
N_{nl} = \left[\frac{1}{z_{nl}^3}\int_0^{z_{nl}} j_l(x)^2x^2\,{\rm d}x\right]^{-1/2}.
\end{equation}
The corresponding Laplacian eigenvalues are 
\begin{equation}
E_{nl} = \frac{z_{nl}^2}{a^2}.
\label{eq:EnlZPE}
\end{equation}

When computing the density expansion coefficients in this basis it is necessary to introduce a cutoff, so that only neighbors within $\rcut=a$ of the central atom contribute to $\rho_i(\bx)$. For this we use modified Gaussian smearing functions of the form
\begin{equation}
\rep<\bx||\br_{ji}; \tilde{g}>\equiv f_1(x) f_2(r_{ji})\exp(-|\bx-\br_{ji}|^2/2\sigma^2),\label{eq:smear}
\end{equation}
in which $f_1(x)$ goes to zero linearly as $x\to a$ so as to fulfil the boundary condition of the LE problem and $f_2(r_{ji})$ goes to zero quadratically as $r_{ji}\to a$ to ensure that properties obtained from the expansion are continuous along with their gradients as atoms enter and leave the sphere. A simple choice for these cutoff functions, which we have used in our calculations on the carbon dataset, is
\begin{equation}
f_k(r) = \Big[\cos\Big(\frac{\pi r}{2a}\Big)\Big]^k.
\end{equation}

The modified Gaussian smearing functions in Eq.~\eqref{eq:smear} can be evaluated in the LE basis as
\begin{equation}
\rep<nlm||\br_{ji}; \tilde{g}> =
\rep<nl||r_{ji}; \tilde{g}> \rep<lm||\brhat_{ji}>.
\end{equation}
The radial integrals are given by
\begin{multline}
\rep<nl||r; \tilde{g}>\equiv \tilde{g}_{nl}(r)=
4\pi f_2(r)\times \\
\int_0^a R_{nl}(x)f_1(x)e^{-(x-r)^2/2\sigma^2}e^{-xr/\sigma^2}i_l(xr/\sigma^2)\,x^2{\rm d}x,
\end{multline}
where $i_l(x)$ is a modified spherical Bessel function.
These radial integrals can be pre-computed on a grid of $r$ values and fit to cubic splines, making the sum over $j$ in Eq.~\eqref{eq:expan-coeffs} cheaper to evaluate for each new arrangement of the atoms. A computer program that implements these equations is provided in the supplementary material.

In the methane tests considered in the text, no hydrogen atom is ever more than $r_{\rm max}=3$ \AA\ away from the central carbon atom, so it suffices to dispense with the cutoff functions and set $a = r_{\rm max}+2.5\sigma$. The cutoff functions should be eliminated for the Jacobian variance tests in any case, so this is how we performed all our methane calculations.

\section{Rayleigh-quotient formulation of optimized bases}
\label{app:rayleigh}

In the main text we have constructed the X-OPT and J-OPT bases by diagonalising the covariance matrices in Eqs.~\eqref{eq:Clambda} and~\eqref{eq:Klambda} in a large primitive basis. It is however also possible to introduce a framework that reveals a closer connection between these data-driven bases and the LE construction. In fact, all orthogonal sets of basis functions can be thought of as solutions to an appropriate eigenvalue problem or, equivalently, as the orthogonal functions that optimize an appropriate Rayleigh quotient. 

The X-OPT basis is designed to focus on regions where the densities $\rho_i$ are usually large, and the J-OPT basis on regions where their gradients are large. To formalize the concept of ``where the density is usually large'', we can introduce the function
\begin{align}
	c(\bx,\bx') = \sum_i \rho_i(\bx) \rho_i(\bx'),
\end{align}
where the sum runs over all atomic environments. This function measures how correlated the densities are at the locations $\bx$ and $\bx'$. Similarly, we can introduce a gradient analogue
\begin{align}
	k(\bx,\bx') = \sum_{i\alpha} \frac{\partial \rho_i}{\partial x_\alpha}(\bx) \frac{\partial \rho_i}{\partial x_\alpha'}(\bx'),
\end{align}
that measures correlations between the derivatives of the neighbor density. Using these functions, we can define the Rayleigh quotients
\begin{align}
	Q_X = \frac{\int \mathrm{d}\bx \int \mathrm{d}\bx' \,u(\bx)c(\bx,\bx') u(\bx')}{\int |u(\bx)|^2\,\mathrm{d}\bx}
\end{align}
and 
\begin{align}
	Q_J & = \frac{\int \mathrm{d}\bx\int \mathrm{d}\bx'\, u(\bx)k(\bx,\bx') u(\bx')}{\int |u(\bx)|^2\,\mathrm{d}\bx},
	\label{eq:r-qj}
\end{align}
that are maximized by the X-OPT and J-OPT bases, respectively. (It is also possible to find linear operators $L$ such that the X-OPT and J-OPT basis functions are solutions to 
\begin{align}
    Lu = \lambda u, 
\end{align}
and to write the associated Rayleigh quotients as
\begin{align}
	Q = \frac{\langle u | L | u \rangle}{\langle u | u \rangle},
\end{align}
as discussed in more detail in the SI.)

A key difference between these bases and the LE basis is that the X-OPT and J-OPT bases {\em maximize} these Rayleigh quotients, whereas the LE basis minimizes the Rayleigh quotient in Eq.~\eqref{eq:rq}. This difference can be used to shed some qualitative light on why it is that the Jacobian condition numbers obtained from the J-OPT basis in Fig.~\ref{fig:carbon-cn} are so large. An integration by parts in Eq.~\eqref{eq:r-qj} gives
\begin{align}
	Q_J & = \frac{\int \mathrm{d}\bx\int \mathrm{d}\bx'\,  c(\bx, \bx')\, \nabla u(\bx)\cdot \nabla u(\bx')}{\int |u(\bx)|^2\,\mathrm{d}\bx}.
\label{eq:r-qj-parts}	
\end{align}
This has a similar form to the Rayleigh quotient in Eq.~\eqref{eq:rq}, but with a non-local (positive semi-definite) kernel $c(\bx,\bx')$ rather than a delta function $\delta(\bx-\bx')$. The LE basis minimizes the Rayleigh quotient in Eq.~\eqref{eq:rq} to give smooth basis functions, whereas the J-OPT basis {\em maximizes} the Rayleigh quotient in Eq.~\eqref{eq:r-qj-parts} to give basis functions that are considerably less smooth.

\section{Iterative variance optimization} \label{app:ivo}

Here we describe a cheaper IVO alternative to the PCA contraction of high $\nu$ equivariant features. This alternative is a variation on a theme of the iterative CUR selection,\cite{imba+18jcp} which proceeds by selecting the feature with highest variance and then orthogonalizing the remaining features to it. In order to ensure that the equivariance of the features is preserved, we orthogonalize separately features with different $\lambda$ (and $\sigma$) character. This can be achieved in practice by treating the different projection components $\mu$ as if they were separate samples. 

Take $\mathbf{X}$ to be the $[\ntrain (2\lambda+1)]\times n_{\rm feature}$ ``flattened'' feature matrix with angular momentum character $\lambda$. Compute the column-wise norm, and select the index $k$ corresponding the largest norm. All remaining column vectors $\mathbf{X}_j$ are then orthogonalized with respect to  column $\mathbf{X}_k$, 
\begin{equation}
\mathbf{X}_j \leftarrow \mathbf{X}_j - \mathbf{X}_k (\mathbf{X}_k\cdot \mathbf{X}_j)/(\mathbf{X}_k\cdot\mathbf{X}_k),
\label{eq:ivo-orthog}
\end{equation}
the column norm is re-computed, and the selection repeated. By choosing a threshold on the residual norm, we can achieve a truncation that mimics a principal component selection, retaining only the features that contribute most to the variance for each equivariant block.

It is important to recognize that the orthogonalization  \emph{removes} duplicate or linearly dependent features. When using the selected features in a regression scheme, this does not entail any loss of information (e.g. the global feature-space reconstruction error, GFRE,\cite{gosc+21mlst} would be zero). However, it does affect our metrics $\ell_X$ and $\ell_J$, because the reduced feature matrix $\tilde{\mathbf{X}}$ that is obtained by assembling the $\tilde{n}_\text{feature}$ highest variance features will have a lower variance than $\mathbf{X}$ when duplicates have been removed. In other words, the discretized coefficients are not a unitary transformation of the real-space neighbor density (i.e. the compression leads to a large feature-space distortion, or GFRD\cite{gosc+21mlst}), which is incompatible with our goal of obtaining features that yield a lossless encoding of the neighbor distribution. 

Fortunately, this is easily remedied by computing a ``weighing matrix'' $\mathbf{W}$ such that
\begin{equation}
{\bf W}{\bf W}^T = \tilde{\bf X}^+{\bf X}{\bf X}^T(\tilde{\bf X}^+)^T,
\end{equation}
where $(\cdot)^+$ denotes the matrix pseudo-inverse. The weighted features can then be obtained as $\tilde{\mathbf{X}}\mathbf{W}$, and these recover the least-distorted compression for a given selection.
To understand why, consider the case in which only two identical features (columns of {\bf X}) are present, and one is removed. Even though no information is lost, the squared norm of the selected features is halved relative to the starting features, affecting both $\ell_X$ and $\ell_J$. If the retained column is multiplied by $\sqrt{2}$, however, the squared norm of the starting features is recovered, and the procedure is lossless for $\ell_X$ and $\ell_J$. The $\mathbf{W}$ matrix generalizes this idea to multiple features with arbitrary linear correlations to the features that have been removed.

\section{High-order Jacobian}
\label{app:hi-jacobian}

Here we provide a simple (if cumbersome) proof that the Jacobian of any high-order set of equivariants built from a complete discretization of the neighbor density is diagonal and has condition number equal to one, i.e. that the result in Eq.~\eqref{eq:highnu-jcn} holds true. 

To clarify the argument, we shall avoid the complications that arise from resolving the parity index $\sigma$, and work with Eq.~\eqref{eq:highnu-jcn} in its unresolved form
\begin{equation}
\sum_{q\lambda\mu}
\rep<\partial_{j\alpha}\frho[\lambda\mu]_i^{\nu}||q>
\rep<q||\partial_{j'\alpha'}\frho[\lambda\mu]_i^{\nu}>=k_i \delta_{jj'}\delta_{\alpha\alpha'}.
\label{eq:j-nu-sum}
\end{equation}
When $\sigma$ is resolved, the final result is the same, but with an additional sum over the parity index in each of the terms in Eq.~\eqref{eq:XX-XJ-JJ}.

In this simplified notation, a generic application of the CG iteration used in Eq.~\eqref{eq:gen-cg-iter} gives
\begin{multline}
\rep<q_1 l_1; q_2 l_2||\partial_{j\alpha}\frho[\lambda\mu]_i^{(\nu_1+\nu_2)}>=\sum_{m_1m_2} \cg{l_1 m_1}{l_2 m_2}{\lambda\mu}
\\ \times\left[\rep<q_1||\partial_{j\alpha}\frho[l_1m_1]_i^{\nu_1}>\rep<q_2||\frho[l_2 m_2]_i^{\nu_2}>+\right.\\
\left.
\rep<q_1||\frho[l_1m_1]_i^{\nu_1}>\rep<q_2||\partial_{j\alpha}\frho[l_2 m_2]_i^{\nu_2}>
\right].
\label{eq:j-cg}
\end{multline}
Combining two terms of this form, and exploiting the orthogonality of CG coefficients
\begin{equation}
\sum_{\lambda\mu} 
\cg{l_1 m_1}{l_2 m_2}{\lambda\mu}
\left<\lambda\mu|l_1m_1';l_2m_2'\right> =\delta_{m_1m_1'}\delta_{m_2m_2'},
\end{equation}
we find that the left-hand side of Eq.~\eqref{eq:j-nu-sum} can be written as the sum of four terms, each of which is associated with a product of two lower-order contractions:
\begin{multline}
\sum_{q\lambda\mu}
\rep<\partial_{j\alpha}\frho[\lambda\mu]_i^{\nu}||q>
\rep<q||\partial_{j'\alpha'}\frho[\lambda\mu]_i^{\nu}> \\
=J_{j\alpha}J_{j'\alpha'}(\nu_1) XX(\nu_2) +
XX(\nu_1) J_{j\alpha}J_{j'\alpha'}(\nu_2) \\
+XJ_{j\alpha}(\nu_1) XJ_{j'\alpha'}(\nu_2) +
XJ_{j\alpha}(\nu_2) XJ_{j'\alpha'}(\nu_1),\\
\end{multline}
where $\nu=\nu_1+\nu_2$ and
\begin{equation}
\begin{split}
XX(\nu_k) =& 
\sum_{qlm} \rep<\frho[lm]_i^{\nu_k}||q>
\rep<q||\frho[lm]_i^{\nu_k}>\\
XJ_{j\alpha}(\nu_k) =& 
\sum_{qlm} \rep<\partial_{j\alpha}\frho[lm]_i^{\nu_k}||q>
\rep<q||\frho[lm]_i^{\nu_k}>\\
J_{j\alpha}J_{j'\alpha'}(\nu_k)=&\sum_{qlm} \rep<\partial_{j\alpha}\frho[lm]_i^{\nu_k}||q>
\rep<q||\partial_{j'\alpha'}\frho[lm]_i^{\nu_k}>.
\end{split}
\label{eq:XX-XJ-JJ}
\end{equation}

In the limit of sharp Gaussians and a complete basis, $XX(1)$ is proportional to the number of neighbours $n_i$ of atom $i$, and is independent of the positions of these neighbours. As a consequence, $XJ_{j\alpha}(1)$ is zero. And $J_{j\alpha}J_{j'\alpha'}(1)$ is just the Jacobian of the density expansion coefficients, which is shown in Ref.~\onlinecite{pozd+21ore} to be a constant times $\delta_{jj'}\delta_{\alpha\alpha'}$. 
This proves~\eqref{eq:j-nu-sum} for $ \nu_1=\nu_2=1$ and $   \nu=2$. Considering that the density sum rule~\eqref{eq:norm-product} implies that $XX(\nu_k)\propto n_i^{\nu_k}$, and therefore $XJ_{j\alpha}(\nu_k)=0$, the general case can be verified by induction.

\section{LE-ACE model}
\label{app:le-ace}
\rev{ The regression results shown in Sec. \ref{sec:LE-ACE} were obtained by incorporating the LE basis into ACE\cite{drau19prb}. In this section, we briefly describe the peculiarities of the resulting LE-ACE model.

As in the original ACE method, delta-like densities are employed and a radial transform is used to give more weight to short-range interactions\cite{drau19prb, kov2021jctc, dusson_2022_atomic}. In our model, the radial transform takes the form
\begin{equation}
x = \xi(r) = a\Big(1-\textrm{exp}\Big(- f \,\, \textrm{tan}\Big(\frac{\pi r}{2a}\Big)\Big)\Big)\Big),
\label{eq:radialtransform}
\end{equation}
where $r$ is the original radial coordinate, $x$ is the transformed coordinate, $a$ is the radius of the sphere inside which the Laplacian eigenstates are defined, and $f$ is a multiplicative factor that can be optimized. This radial transform, unlike others in the literature, can be applied directly to the LE basis. The resulting basis functions $\psi(\xi(r))$ have the desirable property of going to zero, along with all their derivatives up to infinite order, at $r = a$. This guarantees the continuity of the predicted target property and its derivatives of any order at the cutoff radius.

Following Ref. \citenum{kov2021jctc}, whose results are reproduced in Table \ref{tab:LE-ACE} under the ACE label, we use different cutoff radii for the two-body ($\nu = 1$) interactions and for higher body-order interactions. These cutoff radii are set to 5.5 \AA\, and 4.4 \AA, respectively, as they are in Ref. \citenum{kov2021jctc}. It should be noted that these choices for the cutoff radii, while tuned for ACE and also adopted here for consistency purposes, may not be optimal for LE-ACE, which uses different basis functions and a different radial transform.

Tikhonov regularization is employed during the linear fitting stage. Following Ref. \citenum{kov2021jctc}, the regularization matrix is chosen to be diagonal with entries corresponding to estimates of the derivatives of the basis functions. However, while in Ref. \citenum{kov2021jctc} the degree of the estimated derivative is an optimizable parameter, here we always choose to use an estimate of the \textit{first} derivative, given by $\sqrt{E_b}$, where $E_b$ is the eigenvalue of the (possibly high-order) LE basis function $b$. See Section \ref{sec:LEbasis} for more details on the connection between $E_b$ and the mean square gradient of the LE basis functions.

Given the points above, in order to fully define the LE-ACE model, only the maximum correlation order $\nu_{\rm{max}}$ and the maximum Laplacian eigenvalues $E_{\rm{max}}(\nu)$ for $1 \le \nu \le \nu_{\rm{max}}$ need to be specified. $\nu_{\rm{max}}$ is set to 3 for azobenzene, uracil, and paracetamol, and to 4 for all other molecules in the rMD17 dataset. These parameters, which ensure consistency with the reference ACE benchmarks\cite{kov2021jctc}, are almost certainly not optimal. This needs to be taken into consideration when comparing LE-ACE with NequIP and MACE, which can theoretically describe interactions up to infinite body-orders. A more justifiable choice is that of $\nu_{\rm{max}} = 4$ for random methane (where higher-body-order contributions are not relevant). 

Regarding the $E_{\rm{max}}(\nu)$ thresholds, these should be adapted according to the importance of the body-ordered contributions in each specific chemical system and to the extent to which the training set samples the potential energy surface. While the details are relegated to the SI, we emphasize that we did not optimize these parameters separately for each molecule in the rMD17 dataset (as was done in the ACE benchmarks). Instead we used a single set of coefficients for all molecules for which $\nu_{\rm{max}} = 4$, and a different set for those for which $\nu_{\rm{max}} = 3$ (see SI).
}

\onecolumngrid
\clearpage
\newpage

\clearpage
\foreach \x in {1,2,3,4,5,6,7,8,9,10,11,12,13,14}
{%
\clearpage


\section*{Supplementary material}

# Supplemental Information

## 1 Low-energy methane structures

The nearly random nature of the random methane dataset does not leave much scope for optimization with a data-driven basis. In order to test structures where more room for optimization is available, the 10 000 lowest-energy structures were extracted from the 7 732 488 structures in the random methane dataset, and subsequently used as an independent dataset. Figure S1 shows the distribution of the angular tetrahedral order parameter $S_g$ presented in Ref. 34 in the low-energy subset and in a random 10k subset of the random methane dataset.

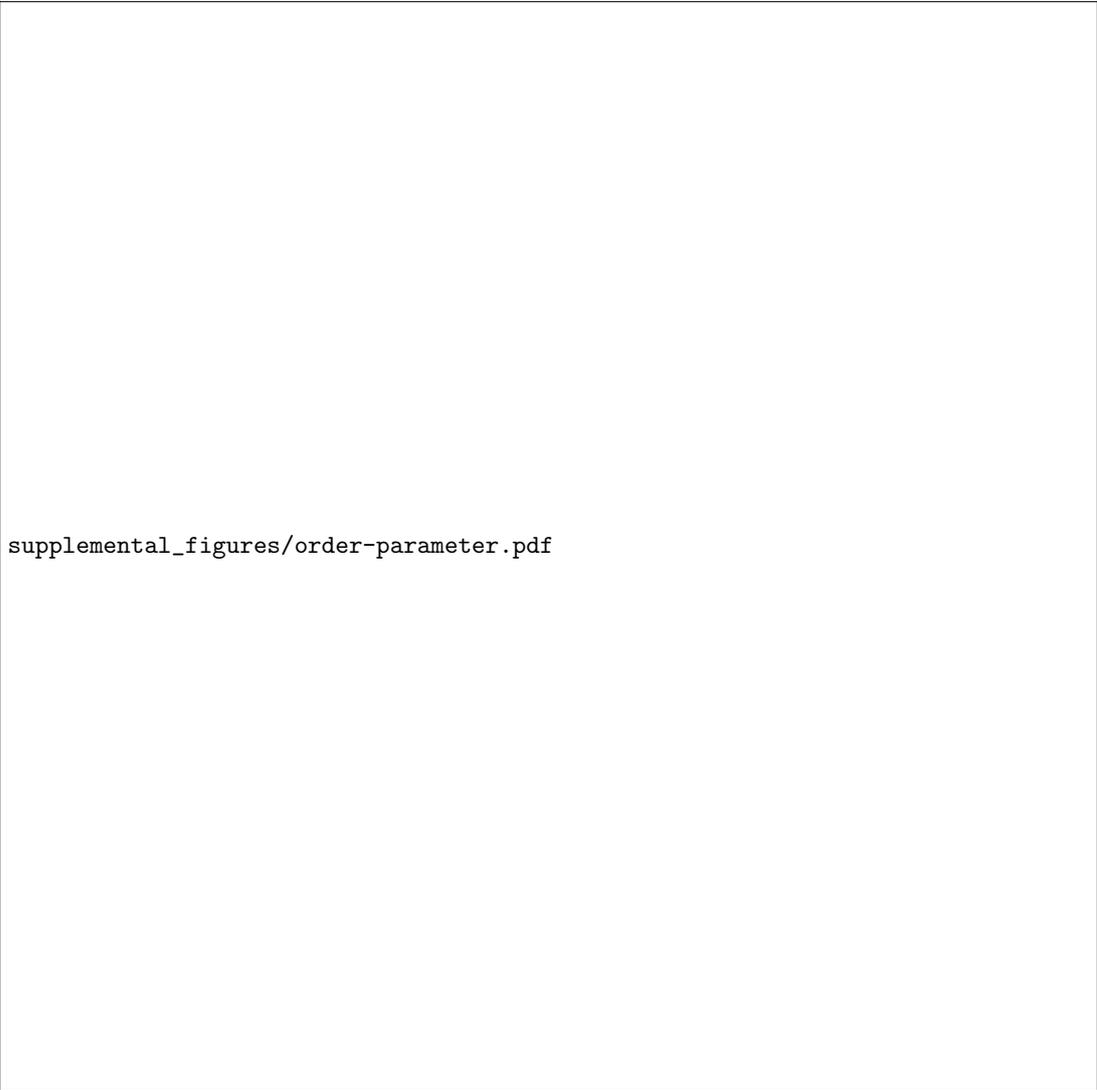

Figure S1: Angular tetrahedral order parameter $S_g$ for 10k random structures in the random methane dataset and for the 10k lowest-energy structures. Each blue dot represents the $S_g$ parameter for a single structure, while the red line represents the average value in the two sub-datasets.

The value of $S_g$ varies from 0 to 1, and it is 0 for a perfect tetrahedron. The average $S_g$ value for the random structures is 0.256823, which is very close to the theoretical average value of $1/4$ for randomly distributed points[34]. In contrast, the average order parameter is 0.188703 for the low-energy structures, indicating that they are, on average, significantly closer to an ideal tetrahedral configuration.

## 2 Full results

This section contains the results of the tests comparing different bases and truncation criteria. In all cases, a LE basis truncated with the "rectangular" $l_{max} = 40$, $n_{max} = 25$ parameters was employed as the large basis from which all data-driven bases were built. This basis was found to be converged in all the performed tests, and it was also used to measure the variance and Jacobian variance of the structures in the complete basis limit. The density smoothing parameter $\sigma$ was set to 0.2 Å both for the methane dataset and for the AIRSS carbon dataset. Larger values result in a plateau of the condition numbers within the range of basis

sizes considered, while smaller values of $\sigma$ delay the convergence of the condition numbers to 1. In all cases, the $a$ parameter of the LE expansion was set to $r_{\text{cut}} + 2.5\sigma$ in order to guarantee the expansion of the vast majority of the density around each atom in the relevant neighborhood. This is an important consideration in the Jacobian condition number tests.

## 2.1 Density expansion coefficients

When comparing the density expansion coefficients across different bases and basis sizes, each basis is truncated by its own natural truncation criterion. Specific $(n_{\max}, l_{\max})$ pairs are also shown for the GTO and X-OPT bases, as this is the way these two bases had been truncated previously to this work. Due to their shape in an $n_{\max} - l_{\max}$ plot (e.g., Fig. 1), these truncations will be referred to as "rectangular". For each test, three rectangular truncation parameter pairs will be shown: $(n_{\max}, l_{\max}) = (6, 4), (8, 6), (12, 9)$. These are represented by isolated square markers in the plots.

### 2.1.1 Variance tests

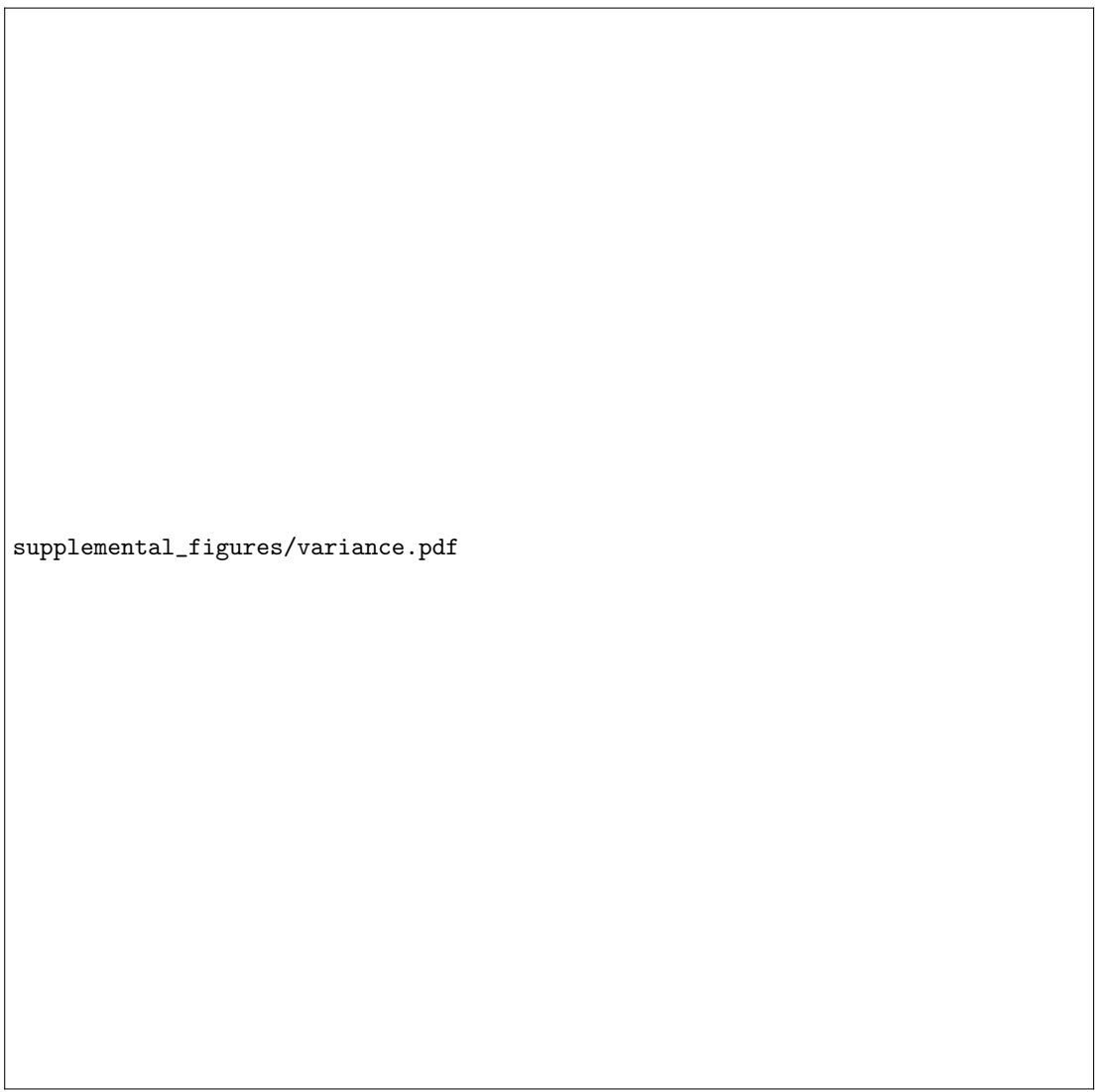

Figure S2: Variance tests.

From Fig. S2, it can be seen that, naturally, all curves are monotonic: as the size of the basis increases, it captures more of the atomic density in the structures. The X-OPT basis achieves the lowest residual variance in all cases, as it is optimized to do so. The rectangular bases suffer from the exclusion of high-$l$ contributions, except in the low-energy methane case, where the tetrahedral nature of the structures favors $l \leq 3$ channels. In contrast, the natural truncation criteria used for the LE, X-OPT and J-OPT restrict the truncation of the basis to a single parameter (which unequivocally specifies the size of the basis), and they outperform the $(n_{max}, l_{max})$ truncations in the vast majority of cases. This phenomenon is not limited to the variance tests, but it also holds true in the other tests considered. In particular, it is useful to note the difference in performance between X-OPT and Rect. X-OPT, which are the same basis truncated in different ways, and the similarity between GTO and Rect. X-OPT in all considered tests, which highlights that a poor truncation choice can almost nullify the efforts made in optimizing the basis functions themselves.

It can be seen that the LE basis fails to optimize for low-energy methane. Indeed, while the other bases capture more variance in the low-energy case, the LE curves are almost exactly the same as for random methane. In this context, it is worth noting that the GTO basis, despite not being data-driven, captures more variance in the low-energy structures because it focuses on the closest regions to the central atom, and this is beneficial when expanding the low-energy densities, where most of the H atoms are close to the C center (relative to the 3 Å cutoff radius). Despite its inability to adapt to the specific datasets, the LE basis is always competitive, and, given an X-OPT basis of a certain size, only a slightly larger LE basis is needed to achieve the same residual variance. Moreover, the fact that the carbon tests reproduce more closely the behavior of the random methane tests - rather than the low-energy ones - indicates that realistic bulk structures display closer to random behavior rather than highly organized low-energy-methane-like distributions.

### 2.1.2 Jacobian variance tests

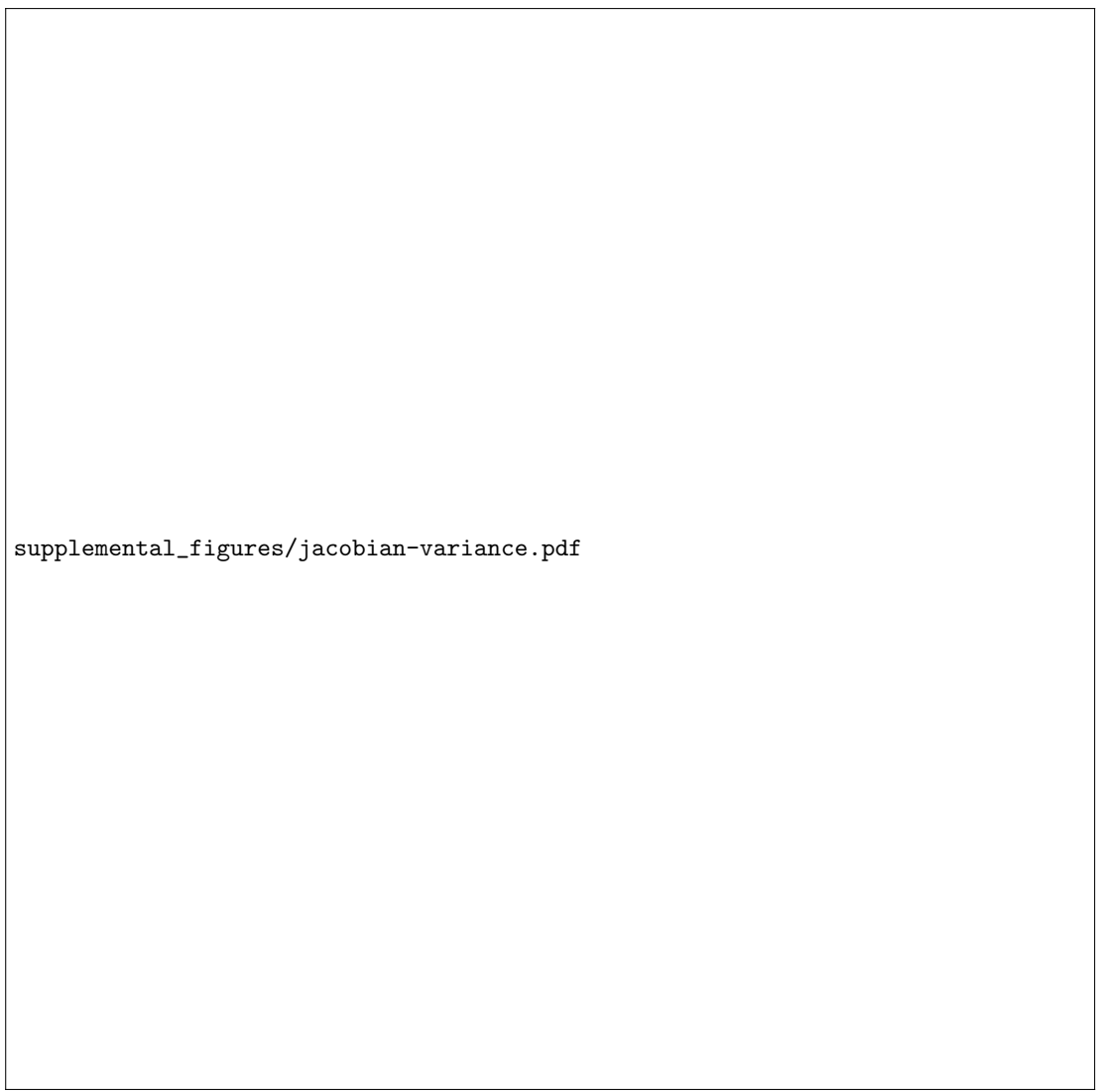

Figure S3: Jacobian variance tests.

Fig. S3 confirms that J-OPT is the optimal basis on the Jacobian variance test (by construction), as it always outperforms all other considered bases. Similarly to the variance tests, and despite the fact that the LE basis cannot adapt to different datasets, it is nonetheless competitive with J-OPT and the other bases in all cases, with the possible exception of the low-energy methane structures. Once again, the rectangular bases suffer from the lack of high-$l$ channels (but not in the low-energy methane case, where high-$l$ contributions are less important due to the high symmetry of the structures). It is also interesting to note how the X-OPT basis performs almost as well as the J-OPT basis in this test, while the opposite is not true (i.e. J-OPT is noticeably worse than X-OPT) in the variance tests. This perhaps suggests that, as an optimized basis, X-OPT is a better choice than J-OPT.

### 2.1.3 Jacobian condition number tests

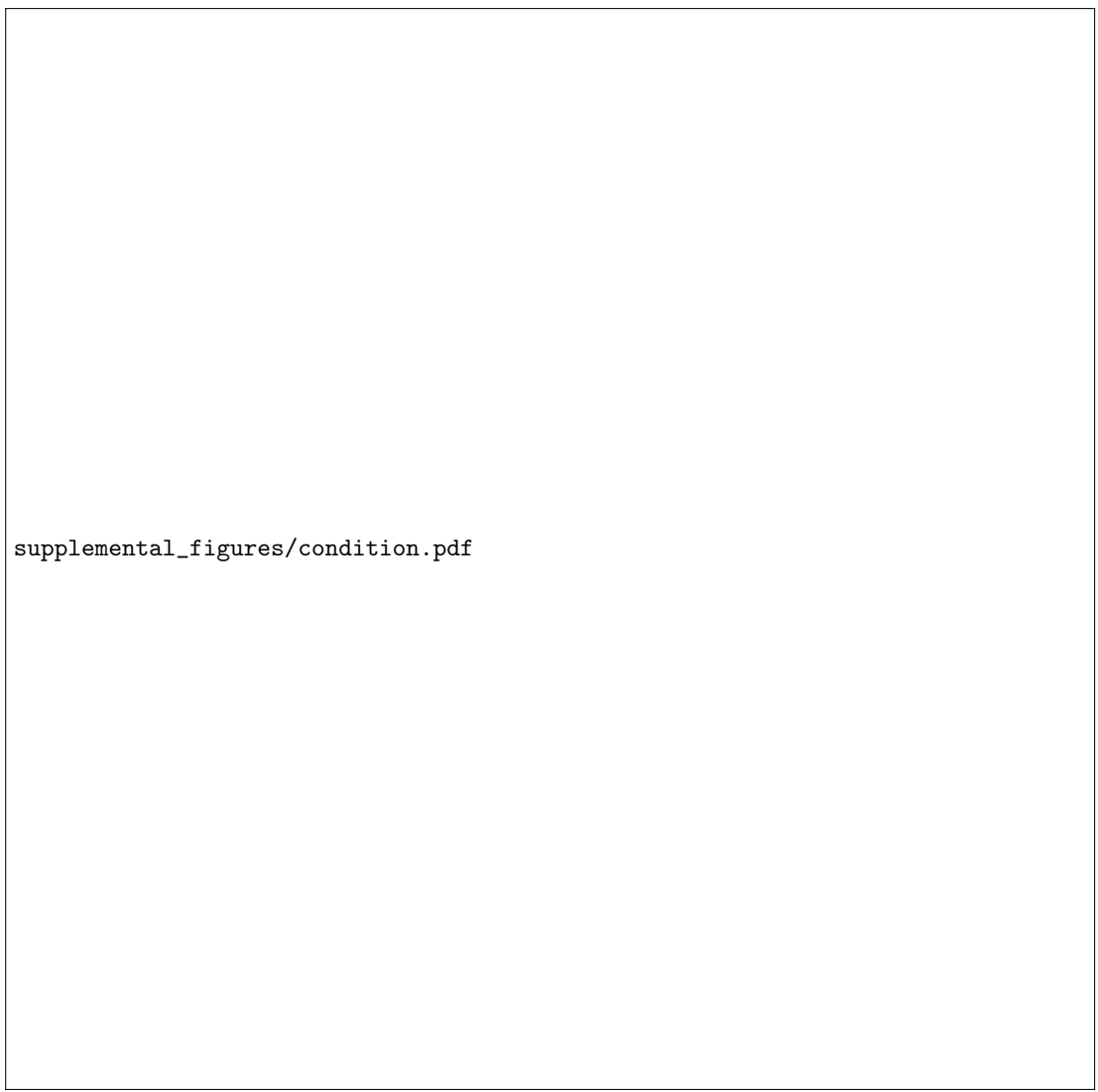

Figure S4: Jacobian condition number (CN) tests.

The Jacobian condition number tests in Fig. S4 reveal that the LE basis leads to a highly uniform and isotropic mapping of real space into feature (expansion coefficient) space. The fact that the data-driven bases perform worse than LE, and especially when the room for optimization is larger (e.g., low-energy methane), follows from the fact that the Jacobian CN is a measure of the distortions caused by the transformation from real space to feature space. The GTO and Rect. X-OPT bases generally perform worse than the others: their arbitrary truncation leads to the exclusion of important spherical expansion components (in particular, the complete exclusion of high-$l$ channels) and the inclusion of less important ones. Specifically, the sharp change in the number of selected basis functions as a function of $l$ (from $n_{max}$ at $l_{max}$ to 0 for $l > l_{max}$) results in a high degree of distortion which translates to large condition numbers. As noted in the previous tests, the two bases truncated via the rectangular truncation criteria (GTO and Rect. X-OPT) achieve essentially the same performance, suggesting that the truncation criterion is a crucial feature of a given basis, and that it is possibly even more important than the choice of the basis functions.

## 2.2 Equivariant $\nu = 2$ features

Here, and in the $\nu = 3$ learning exercises, only the LE and X-OPT bases are taken into consideration, as they both perform well in our $\nu = 2$ learning exercises and generalize in a consistent way to higher body-orders. The generalizations of the LE and X-OPT bases are, respectively, the high-order LE selection and PCA-like contractions of the high-order equivariants, as detailed in the main text. IVO (presented in Appendix C) will also be tested as a computationally cheaper alternative to PCA.

All parameters in the density expansions were kept the same as in the $\nu = 1$ tests above. For each LE-LE equivariant set size in the following plots, the $\nu = 1$ LE basis was chosen as the smallest possible basis that enables the computation of all required LE-LE $\nu = 2$ equivariants (i.e., the $\nu = 1$ LE energy threshold is found from Eq. 29). Starting from this "minimal" $\nu = 1$ LE basis, all the available $\nu = 2$ equivariants were calculated (including those above the $\nu = 2$ LE threshold). IVO and PCA were then used to contract the $\nu = 2$ equivariant set to a similar size as the corresponding LE-LE equivariant set. These two contractions constitute the LE-IVO and LE-PCA $\nu = 2$ bases. Finally, a $\nu = 1$ X-OPT basis was calculated in the same manner as for the $\nu = 1$ methane tests. The size of the X-OPT basis was chosen so as to be similar to that of the corresponding "minimal" $\nu = 1$ LE basis. Starting from this X-OPT basis, all available $\nu = 2$ equivariants were calculated. The X-OPT-IVO and X-OPT-PCA $\nu = 2$ bases were then obtained by performing IVO and PCA on the resulting equivariant set to yield equivariant sets of similar size as LE-LE. [In this context, a "similar" equivariant count corresponds to a tolerance of $\pm 25\%$.]

### 2.2.1 Variance tests

The variance of the $\nu = 2$ equivariants is calculated in the same way as that of the $\nu = 1$ equivariants (Eq. 12).

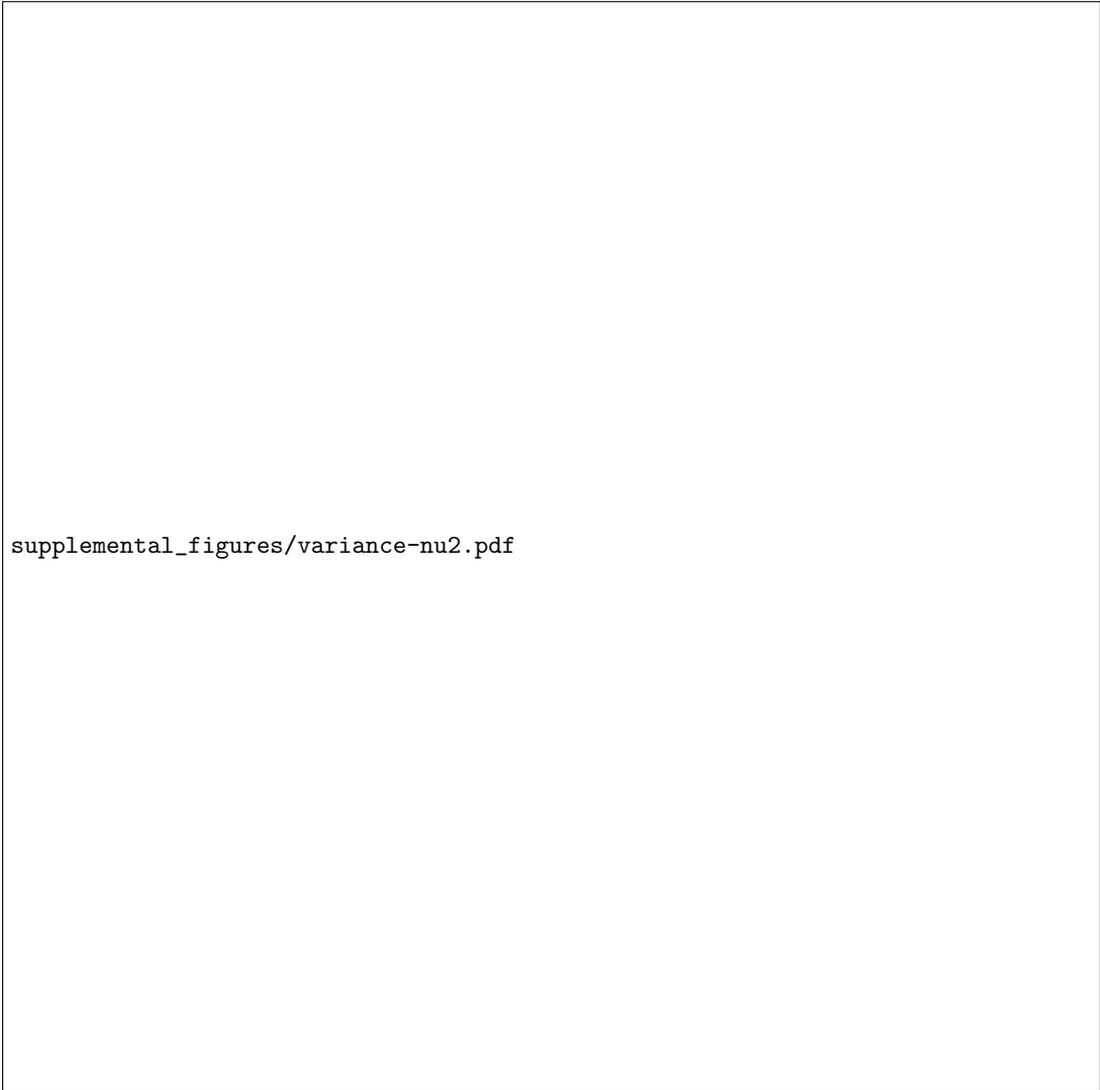

Figure S5: $\nu = 2$ variance tests.

Fig. S5 shows that, although the $\nu = 2$ contractions do have an effect on the $\nu = 2$ variance tests, the majority of the performance differences arise from the $\nu = 1$ bases, at least with our "minimalist" $\nu = 1$ truncation approach. Indeed, it should be noted that LE-PCA and X-OPT-PCA would converge to the same basis in the limit of complete $\nu = 1$ expansions. Our choice, however, is closer to the scenario of a real application, a case in which minimizing storage requirements and computational cost would be desired. As noted in the $\nu = 1$ tests, all bases perform similarly on the random methane dataset, while their performances differ more significantly on the low-energy methane structures. Unsurprisingly, X-OPT-PCA performs better than the other truncations on both datasets. However, the difference in performance between the latter and LE-LE is larger than in the $\nu = 1$ case, and this is especially true for low-energy methane. This hints to the $\nu = 2$ features being much more compressible than the $\nu = 1$ features, a fact that can be rationalized by considering that a full set of $\nu = 2$ features contains the same information as that of the starting $\nu = 1$ equivariant set, while its size is formally its square. The IVO selection scheme affords close to optimal (PCA) performance, indicating a high degree of orthogonality (within the dataset) of $\nu = 2$ features with the same symmetry. With IVO, once the selected $\nu = 2$ features are determined on a training set (which consists of the whole dataset in these tests, but not in the learning exercises), the IVO contraction allows for the computation of the retained features *only*, similarly to the LE high-order truncation, and unlike PCA, where

all equivariants need to be computed, and stored, before the contraction step.

### 2.2.2 Jacobian condition number tests

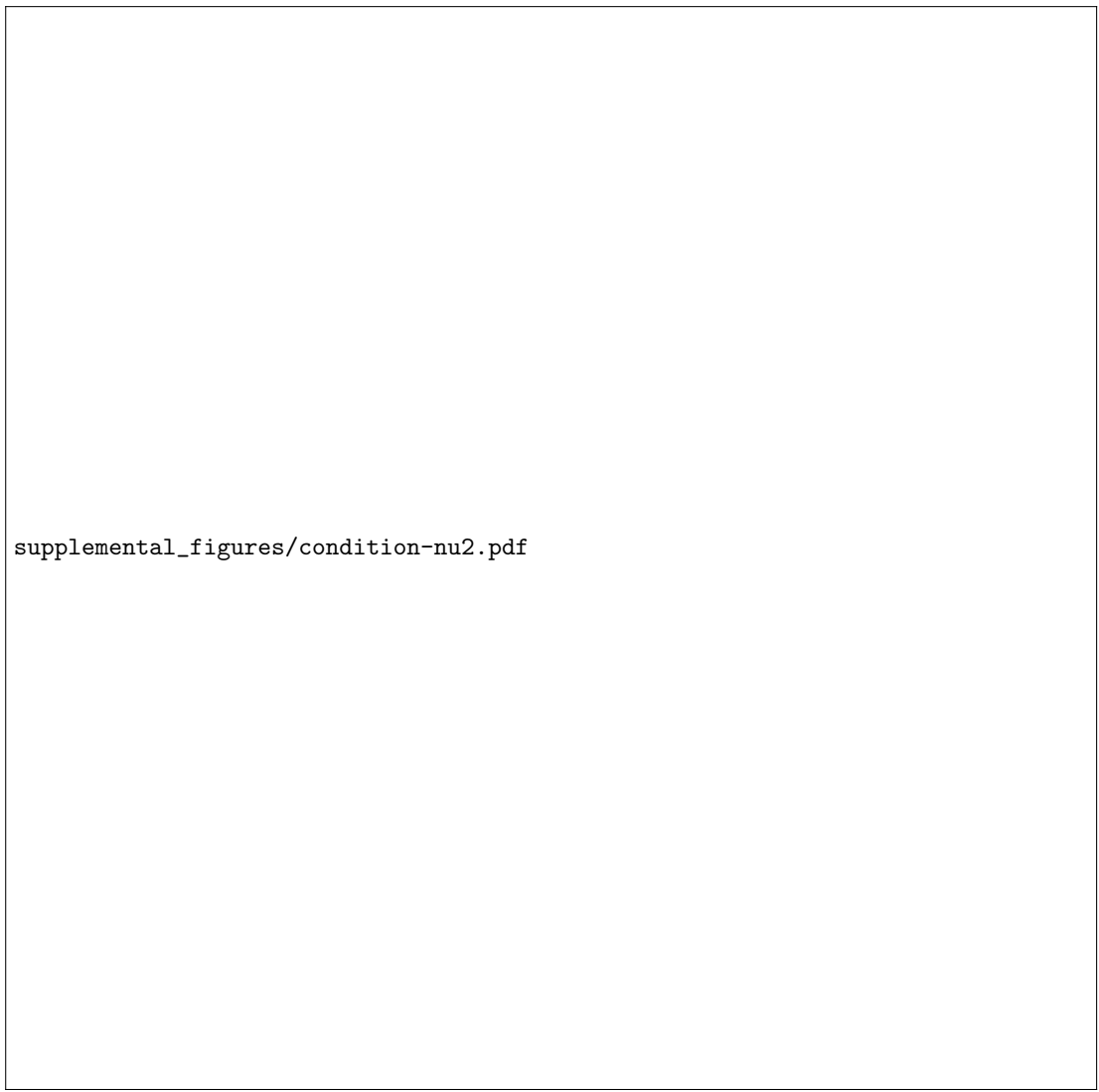

Figure S6: $\nu = 2$ Jacobian condition number tests.

The $\nu = 2$ Jacobian condition number tests in Fig. S6 show that, similarly to the $\nu = 2$ variance case, the choice of $\nu = 2$ compression has little consequences compared to the $\nu = 1$ basis, possibly as a result of the strict $\nu = 1$ thresholds, which leave relatively little scope for differences in the $\nu = 2$ selection/contraction. It is also worth noting that the X-OPT-PCA and X-OPT-IVO contractions perform poorly for the low-energy methane structures, confirming that aggressive optimization of a feature set has a negative effect on the Jacobian condition number, which is a measure of the distortion introduced in the transformation from real space to feature space. This is particularly evident from the very high condition numbers generated by the X-OPT-IVO and X-OPT-PCA combinations on low-energy methane.

## 2.3 Learning: up to $\nu = 2$ invariants

In all cases, the learning target was the potential energy of the structures. Each $\nu = 1$ basis was employed in three sizes ("small", "medium" and "large"), truncated with the aim of achieving a very similar number of invariants to that obtained from the three $(n_{\max}, l_{\max})$ truncation parameter pairs mentioned in Sec. 2.1 for GTO (and R-X-OPT). While impossible to match exactly, the number of invariants within bases of the same size was kept nearly the same for all different bases and truncations. For each basis size, all $\nu = 1$ and $\nu = 2$ invariants computable from the density expansion were employed, and the central atoms were not included in the density expansions. The adaptive bases were only allowed to learn from the training set. The learning was performed with linear regression, kernel ridge regression and neural networks in order to demonstrate that the relative performance of the different bases is mostly consistent across different fitting models. For the methane dataset, only carbon-centered features were employed in the learning process, and the feature-related hyperparameters were kept the same as those employed in the expansion coefficient tests. The test sets consist of 2000 structures for the methane sub-datasets and 2931 structures for the carbon dataset.

When learning from the carbon dataset, $r_{cut} = a = 5$ Å was chosen, and a shifted cosine cutoff function (0.5 Å wide) was applied in all cases if not otherwise specified. It must be noted, however, that this choice is not optimal for the LE basis and the optimized bases calculated from it, because the LE basis cannot expand finite densities at the edge of the sphere of radius $a$, where all of its basis functions are zero (Dirichlet boundary condition). As a result of this, and as a means of exploring the effect of more aggressive cutoff functions, the LE calculations were repeated with the $f_1$ and $f_2$ cutoff functions described in Appendix A. Similarly, the X-OPT basis was re-calculated starting from these new LE expansion coefficients. Finally, the GTO results were repeated with a shifted cosine cutoff as wide as $r_{cut}$, which is equivalent to $f_2$.

**Linear regression** The LAPACK dgelsd subroutine was used to perform linear fitting in all instances. Within this subroutine, regularization is performed by excluding the singular values of the feature matrix below a certain threshold given by the `rcond` parameter times the largest singular value. The optimal `rcond` was found via a logarithmic one-dimensional grid search with the aim of minimizing the test set error.

**Kernel ridge regression** Full Gaussian process regression (GPR) was employed with the polynomial kernel $k(\mathbf{x}_i, \mathbf{x}_j) = (\mathbf{x}_i \cdot \mathbf{x}_j)^2$, and the only scaling previous to the fitting is that of the mean potential energy, which was set to zero. The optimal Tikhonov regularization parameter was found via a logarithmic one-dimensional grid search with the aim of minimizing the test set error.

**Neural Networks** In all cases, the feed-forward MLP-type architecture was used. The NNs employed were three layers deep, with each layer consisting of 16 neurons, and tanh activation functions. When learning from the AIRSS carbon dataset, a Behler-Parrinello construction with the same MLP architecture was used. Training of the neural networks was carried out with the Adam optimizer, starting from a $10^{-3}$ learning rate and decreasing learning rates by a factor of 10 after a patience of 50 epochs upon stagnation of the test set error. The optimal point of the learning was either determined by the early-stopping technique or at 10000 training epochs, whichever occurred first. For each point in the learning curves, in order to reduce noise, the results were averaged over 16 training runs starting from randomly initialized weights.

**Results** Fig. S7 shows the full results of the learning exercises using up to $\nu = 2$ invariants. It can be seen that linear models saturate within the $n_{\text{train}}$ range considered due to their inability to represent energy contributions stemming from four- or more body interactions. Conversely, kernel and NN models do not saturate, and they achieve lower RMSEs. While neither is specifically optimized, the performance of NNs and kernels is similar in this context, with GPR having a slight edge for small training sets and NNs achieving slightly lower RMSEs with large training sets. In most cases, the small-sized bases considered here do not include enough features to fit the potential energies optimally, while the medium-sized bases are closer to the optimal resolution. Large bases show signs of overfitting for methane, and they do not improve over the medium-sized bases for carbon.

LE is the best basis when considering learning from random and low-energy methane, and it achieves small RMSEs even in its small version. In contrast, GTO clearly outperforms the other bases in the carbon

learning exercises. The fundamental difference between the methane and carbon datasets in this setting is that, while the methane structures only include one coordination shell, the carbon structures with $r_{cut} = 5$ Å include at least three. It is therefore beneficial for the success of the carbon learning exercises to prioritize the closest regions of each atomic neighborhood, which has a more sizeable effect on the corresponding atomic energy as opposed to further regions. All bases except for the GTO basis treat the whole sphere of radius $a$ equally, while the GTO basis functions sample the region closer to the center of the sphere more densely, resulting in better potential energy fits. It should however be noted how the LE basis is still the best among the bases other than GTO, which treat the density in the whole sphere equally. This, and fact that the LE basis outperforms all other bases on random and low-energy methane, suggests that basis function smoothness is a desirable property to achieve low RMSEs.

Hence, for larger systems, where the cutoff radius includes more coordination shells, taking into account the mostly short-ranged nature of the potential energy is also crucial. On the carbon dataset, the inclusion of more aggressive cutoff functions indeed leads to a marked improvement in the RMSEs. In this case, all the aggressively truncated bases (LE, X-OPT, GTO) perform roughly equally well. This is true despite the fact that GTO, but now also X-OPT (due to its density modulation) sample the closer regions more densely than LE, which, however, is likely to compensate due to its higher smoothness.

Contrary to the variance, Jacobian variance, and Jacobian CN tests, truncation of the basis does not seem to have a large impact on the accuracies achieved within the context of these supervised learning exercises. In particular, it is useful to compare X-OPT and R-X-OPT, which consist of the same basis functions truncated in different ways. Indeed, these two show overall similar performance in Fig. S7, with X-OPT performing slightly better on methane (as perhaps expected) and R-X-OPT being more successful on carbon. Once again, this may be explained by the short-ranged nature of the potential: R-X-OPT ignores high-$l$ contributions and focuses on small-$l$ angular channels, leading to partial cancellation of the atomic contributions of the outer coordination shells in each atomic neighborhood and prioritization of the closer regions.

The J-OPT basis shows the worst performance in most cases. This could be related to the very high condition numbers afforded by the J-OPT basis, which are rationalized in Appendix B.

## 2.4 Learning: up to $\nu = 3$ invariants

We also tested the performance of LE and X-OPT (and their generalizations) on linear learning exercises where $\nu = 3$ invariants were also included. For each basis set type, three invariant set sizes were tested: small (around $10^{2.5}$ invariants), medium ($10^3$), and large ($10^{3.5}$). The invariant set sizes could not be matched exactly to these numbers; hence, a 10% tolerance is allowed in all cases. For each invariant set size and for each $\nu = 2$ truncation type (LE-LE, LE-IVO, LE-PCA, X-OPT-IVO, X-OPT-PCA), a $\nu = 2$ equivariant set is found such that, if combined with its corresponding $\nu = 1$ equivariant set (whose truncation is performed in the fashion described for the $\nu = 2$ equivariant tests), the size of the set of all calculated invariants (with $\nu = 0, 1, 2, 3$) falls within the tolerance intervals described above. It is important to note that this truncation strategy (based on Eq. 29 for $\nu = 1, 2$ and with no truncation for $\nu = 3$ other than that provided by the previous body orders) is the most practically convenient and consistent across all truncation types, but it may not be the optimal choice to obtain low RMSEs.

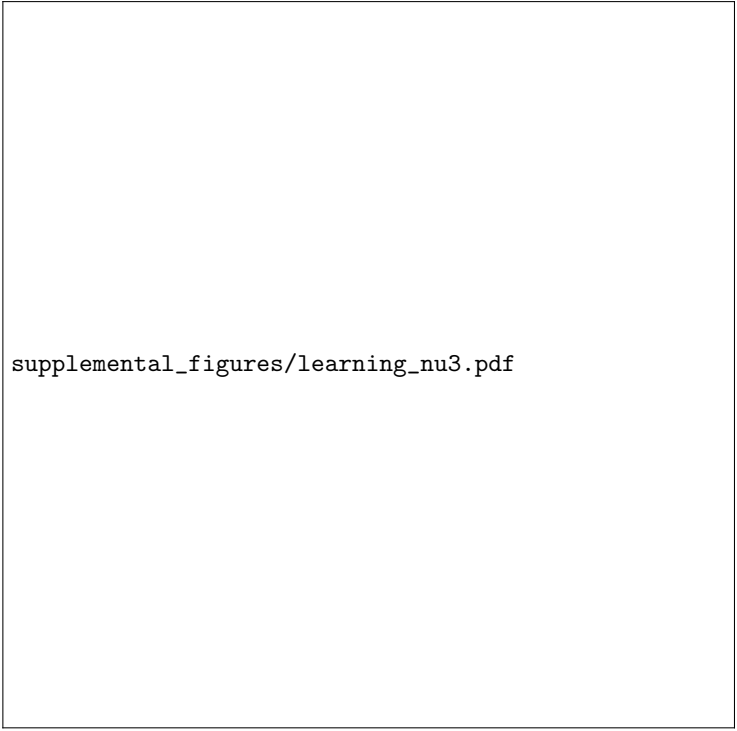

Figure S8: Full results of the learning tests using up to $\nu = 3$ invariants. (s), (m), and (l) denote a small, medium, and large set of $\nu = 3$ invariants, respectively.

Fig. S8 displays the results of the linear learning exercises using up to $\nu = 3$ invariants. As also noted in the $\nu = 2$ variance and Jacobian condition number tests, our minimalist choice for the $\nu = 1$ truncation (for a given $\nu = 2$ equivariant feature set size) leads to only minor differences between the $\nu = 2$ truncation types. Indeed, the RMSE differences are largely accounted for by the $\nu = 1$ LE/X-OPT comparison, which shows the LE basis yielding significantly lower RMSEs, most likely as a result of its smoothness.

Regarding the effect of different $\nu = 2$ contraction schemes, it can be seen that IVO performs as well as or better than PCA in all cases. While these tests do not show large differences between LE-LE, LE-IVO, and LE-PCA, LE-LE performs slightly better on low-energy methane, a dataset where LE-IVO and LE-PCA deviate more from LE-LE due to the non-uniform distribution of the hydrogen atoms. Moreover, partial results with a non-minimalist choice for the $\nu = 1$ bases, where LE-IVO and LE-PCA are allowed to differentiate themselves more significantly from LE-LE, clearly indicated the superiority of LE-LE in terms of RMSEs. Since the LE-LE combination also provides the cheapest way to select high-$\nu$ features, this is clearly the best way to use the LE basis for the machine learning of potential energies.

## 3 A unified picture of optimal basis functions

In this section, we present a more unifying framework for the different optimal bases that are compared in the main text, namely the LE, X-OPT and J-OPT bases. On one hand, all sets of basis functions can be thought of as solutions to an appropriate eigenvalue problem. On the other hand, all three can be defined as the smoothest set of basis functions with a different metric of smoothness. The two viewpoints are closely related to one another, and we shall explore those in the first two subsections. Some mathematical details related to the specific implementation will be addressed in the remaining subsection.

## 3.1 The Eigenvalue picture

The first perspective is the generalization of the eigenvalue problem, where the LE basis functions are defined as solutions to

$$\Delta u = \lambda u. \tag{S1}$$

If we are restricted to a finite number of basis functions, the eigenfunctions corresponding to the largest eigenvalues (rather than the smallest, since we are omitting the minus sign in this version) are selected.
Both the X-OPT and J-OPT bases can be defined similarly as solutions of an eigenvalue problem of the form

$$Lu = \lambda u, \tag{S2}$$

with suitable linear operators $L$. As for the Laplacian, the eigenfunctions corresponding to the smallest eigenvalues are selected here as well.
To provide a short summary for readers fluent in the bra-ket notation, the operators for the X-OPT and J-OPT bases are explicitly given as

$$C = \sum_i |\rho_i\rangle \langle \rho_i| \tag{S3}$$

$$K = \sum_{\alpha=1,2,3} \sum_i |\partial_\alpha \rho_i\rangle \langle \partial_\alpha \rho_i|, \tag{S4}$$

where the sum runs over all atomic environments, i.e. all atoms including every structure in the data set. To make this more explicit, recall that the atomic densities $\rho_i(x)$ around a center atom $i$ are just functions defined on $\mathbb{R}^3$, or to be more precise and taking the finite cutoff radius into account, the 3-dimensional ball of radius $r_{\text{cut}}$

$$\Omega = B^3_{r_{\text{cut}}} = \{\mathbf{x} \in |\mathbf{x}| \leq r_{\text{cut}}\} \subset \mathbb{R}^3. \tag{S5}$$

Using these densities, we can define two new functions (one each for the X-OPT and J-OPT bases) that depend on two variables $\mathbf{x}, \mathbf{x}' \in \Omega$ as

$$c(\mathbf{x}, \mathbf{x}') = \sum_i \rho_i(\mathbf{x})\rho_i(\mathbf{x}'), \tag{S6}$$

where the sum again runs over all atomic environments, and the "gradient version":

$$k(\mathbf{x}, \mathbf{x}') = \sum_{\alpha=1,2,3} \sum_{i=1}^{N_{\text{env}}} \frac{\partial \rho_i}{\partial x_\alpha}(\mathbf{x}) \frac{\partial \rho_i}{\partial x'_\alpha}(\mathbf{x}'). \tag{S7}$$

These functions in two variables can be used to define the two operators $C$ and $K$ as

$$(Cf)(\mathbf{x}) = \int_\Omega c(\mathbf{x}, \mathbf{x}') f(\mathbf{x}') \mathrm{d}x' \tag{S8}$$

$$(Kf)(\mathbf{x}) = \int_\Omega k(\mathbf{x}, \mathbf{x}') f(\mathbf{x}') \mathrm{d}x' \tag{S9}$$

Using these, the first few LE, X-OPT and J-OPT basis functions can be defined as the eigenfunctions of the operators $L_{\text{LE}} = \Delta$, $L_X = C$ and $L_J = K$ with the largest eigenvalues, respectively. Note that $C$ is self-adjoint, while $K$ is so too provided that the densities vanish at the boundaries of the ball.

## 3.2 Smoothest Function Picture

A different way to define the Laplacian eigenstate basis was to consider the Rayleigh quotient

$$Q = \frac{\int_\Omega |\nabla u(\mathbf{x})|^2 \mathrm{d}\mathbf{x}}{\int_\Omega |u(\mathbf{x})|^2 \mathrm{d}\mathbf{x}}. \tag{S10}$$

The basis functions used in the LE basis, can be defined as the functions minimizing this quotient. This is an alternative method that allows us to see the LE basis functions as the smoothest (as in, varying the least) functions we could choose on which to expand the target density.

At the conceptual level, one might imagine that instead of maximizing the overall smoothness over the entire domain, it might make more sense to let the basis functions focus more on regions in which the target densities $\rho_i$ are actually present. This is precisely the idea behind the Rayleigh quotient formulation of the X-OPT and J-OPT bases, which together with the LE basis can all be understood as functions maximizing (rather than minimizing, for convenience) a suitable Rayleigh quotient of the form

$$Q = \frac{\langle u|L|u\rangle}{\langle u|u\rangle}, \tag{S11}$$

where

$$\langle u|u\rangle = \int_\Omega |u(\mathbf{x})|^2 \mathrm{d}\mathbf{x} \tag{S12}$$

is a more compact form to write the denominator in the quotient and the operator $L$ is chosen as in the previous subsection as $L_{\mathrm{LE}} = \Delta, L_X = C$ and $L_J = K$ for the LE, X-OPT and J-OPT bases, respectively.

At the intuitive level, choosing the basis functions maximizing the Rayleigh quotient using the operator $C$ means to pick the functions that have the maximal descriptive capability where the densities are significant. An analogous interpretation can be made for the J-OPT basis using the operator $K$. In the rest of this subsection, we try to provide some further insight into this picture.

We begin by showing how the Rayleigh quotient introduced in the main text can be brought to this form. The connection to the eigenvalue picture can be seen by explicitly writing out

$$|\nabla u(\mathbf{x})|^2 = (\partial_x u)^2 + (\partial_y u)^2 + (\partial_z u)^2 \tag{S13}$$

and integrating the three terms by parts, which allows us to write the numerator as

$$\int_\Omega |\nabla u(\mathbf{x})|^2 \mathrm{d}\mathbf{x} = \int_\Omega u(\mathbf{x}) \cdot (-\Delta u(\mathbf{x})) \, \mathrm{d}\mathbf{x} \tag{S14}$$

$$= \langle u|-\Delta|u\rangle. \tag{S15}$$

Resulting in the previously introduced Rayleigh quotient (with inverted sign for convenience)

$$Q_{\mathrm{LE}} = -\frac{\int_\Omega |\nabla u(\mathbf{x})|^2 \mathrm{d}\mathbf{x}}{\int_\Omega |u(\mathbf{x})|^2 \mathrm{d}\mathbf{x}} = +\frac{\langle u|\Delta|u\rangle}{\langle u|u\rangle}. \tag{S16}$$

Similarly, the X-OPT and J-OPT bases can be interpreted as functions maximizing

$$Q_{\mathrm{X}} = \frac{\langle u|C|u\rangle}{\langle u|u\rangle} \tag{S17}$$

and

$$Q_{\mathrm{J}} = \frac{\langle u|K|u\rangle}{\langle u|u\rangle}. \tag{S18}$$

The interpretation is that these are the functions capturing the maximal amount of the total variation in the atomic densities $\rho_i$ (X-OPT) or their gradients (J-OPT), respectively. Explicitly writing out the integrals, the X-OPT Rayleigh quotient becomes

$$Q_X = \frac{1}{\langle u|u\rangle} \int \mathrm{d}\mathbf{x}\mathrm{d}\mathbf{x}' u(\mathbf{x}) c(\mathbf{x}, \mathbf{x}') u(\mathbf{x}'), \tag{S19}$$

while the J-OPT Rayleigh quotient can be rewritten as

$$Q_J = \frac{1}{\langle u|u\rangle} \int \mathrm{d}\mathbf{x}\mathrm{d}\mathbf{x}' u(\mathbf{x}) k(\mathbf{x}, \mathbf{x}') u(\mathbf{x}') \tag{S20}$$

$$= \frac{1}{\langle u|u\rangle} \int \mathrm{d}\mathbf{x}\mathrm{d}\mathbf{x}' \left(\nabla u(\mathbf{x})\right) c(\mathbf{x}, \mathbf{x}'). \left(\nabla u(\mathbf{x}')\right) \tag{S21}$$


}

\end{document}